\documentclass[fleqn,12pt]{article}
\usepackage{graphics,epsfig,float,mathrsfs,cite,wasysym}

\usepackage{subfigure}

\newcommand{\be}[1]{\begin{equation} \label{(#1)}}
\newcommand{\ee}{\end{equation}}
\newcommand{\baq}[1]{\begin{eqnarray} \label{(#1)}}
\newcommand{\eaq}{\end{eqnarray}}

\newcommand{\rf}[1]{(\ref{(#1)})}

\def\ti  {\tilde}
\newcommand{\CH}{\tilde\chi}

\def\lsim{\raise0.3ex\hbox{$\;<$\kern-0.75em\raise-1.1ex\hbox{$\sim\;$}}}
\def\gsim{\raise0.3ex\hbox{$\;>$\kern-0.75em\raise-1.1ex\hbox{$\sim\;$}}}

\def\Ta{{\mathcal O}_T }
\def\Tb{{\mathcal {\widehat O}}_T }
\def\Tc{{\mathcal {\widehat O}}_T^\prime}
\def\Oa{\langle \Ta \rangle}
\def\Ob{\langle \Tb \rangle}
\def\Oc{\langle \Tc \rangle}
\def\Aa{{\mathcal A}_T }
\def\Ab{\hat{\mathcal A}_T }
\def\Ac{\hat{\mathcal A}_T^\prime}
\def\SOa{S[ \Ta ]}
\def\SOb{S[ \Tb ]}
\def\SOc{S[ \Tc ]}
\def\SAa{S[ \Aa ]}
\def\SAb{S[ \Ab ]}
\def\SAc{S[ \Ac ]}

\newcommand{\AddrVienna}{
\it  Faculty of Physics, Universit\"at Wien, \\ 
A-1090 Vienna, Austria \\}

\newcommand{\AddrCFTP}{
\it Departamento de F\'isica and CFTP, \\
Instituto Superior T\'ecnico Av. Rovisco Pais 1, \\ 
1049-001 Lisboa, Portugal \\}

\newcommand{\AddrGran}{%
\it Departamento de F\'isica Te\'orica y del Cosmos and CAFPE,\\
Universidad de Granada, 
E-18071 Granada, Spain\\}

\newcommand{\AddrHam}{%
\it Institut f\"ur Experimentalphysik, Universit\"at Hamburg\\
Notkestra\ss e 85, D-22607 Hamburg, Germany\\}

\begin{document}

\begin{flushright}
 CFTP/09-021
\end{flushright}

\begin{center}  
  \textbf{\large CP-sensitive spin-spin correlations 
in neutralino production at the ILC}\\[10mm]

{A. Bartl${}^1$, K.~Hohenwarter-Sodek${}^1$, 
T. Kernreiter${}^2$, O.~Kittel${}^3$ and M.~Terwort${}^4$
} 
\vspace{0.3cm}\\ 
$^1$ \AddrVienna
$^2$ \AddrCFTP
$^3$ \AddrGran
$^4$ \AddrHam
\end{center}

\noindent

\begin{abstract}
\noindent 
 We study the CP-violating terms of the spin-spin correlations in neutralino
 production
 and their subsequent two-body decays into sleptons plus leptons at the ILC. 
 We analyze CP-sensitive observables 
 with the help of T-odd products of the spin-spin terms.
 These terms depend on the polarizations of both neutralinos,
 with one polarization perpendicular to the production plane. 
 We present a detailed numerical study of the CP-sensitive observables,
 cross sections, and neutralino branching ratios
 in the Minimal Supersymmetric Standard Model with complex parameters. 

\end{abstract}

\newpage 

\section{Introduction}

It has been pointed out that the amount of CP violation
in the Standard Model (SM) is not sufficient to explain 
the baryon-antibaryon asymmetry of the universe~\cite{Gavela:1993ts}, and
that additional sources of CP violation are required~\cite{Riotto:1998bt}.
Many extensions of the SM can give rise to such sources of 
CP violation. The violation of the CP symmetry is an interesting topic
in its own right and deserves a diligent consideration.
Supersymmetric (SUSY) extensions of the SM provide new sources
of CP violation, as they include several new parameters which
can be complex. 
For instance, in the neutralino sector of 
the Minimal Supersymmetric Standard Model (MSSM) two complex parameters
appear, which lead to CP-violating effects in reactions involving neutralinos. 
These parameters are the higgsino mass parameter $\mu=|\mu|e^{i\phi_{\mu}}$, and 
the U(1) gaugino mass parameter $M_1=|M_1|e^{i\phi_1}$, given in the usual
parametrization of modulus and phase.

These phases, on the other hand, contribute to the electric dipole moments
(EDMs) of electron, neutron, and that of the atoms ${}^{199}$Hg 
and ${}^{205}$Tl~\cite{Ellis:2008zy}, and it
is found in general that for phases of the size ${\mathcal O}(1)$, the 
EDMs are beyond their experimental upper bounds. 
However, the extent to which the EDMs can constrain the CP phases
also depends on most of the other model parameters, and thus
strongly depends on the considered model, see e.g. Refs.~\cite{Ellis:2008zy,EDM}.

In this respect the high-luminosity $e^+e^-$ International Linear Collider~(ILC) 
is considered an ideal machine to perform precision measurements, in order
to determine the model parameters of the MSSM with the required accuracy~\cite{ILC}. 
In neutralino production and decay at the ILC, it has been shown which CP-even 
observables are well suited to access the CP-violating MSSM 
parameters~\cite{Moortgat-Pick:1999di,Kneur:1999nx}. However to directly prove 
CP violation in the MSSM, and to determine the CP-violating phases unambiguously,
a measurement of CP-odd observables is obligatory.

In this paper, we study CP-sensitive observables in neutralino production,
\be{eq:production}
e^+ + e^-\to \CH^0_i + \CH^0_j~,\qquad i,j=2,3,4~, \qquad i \neq j~,
\ee
%
based on T-odd correlations~\cite{Valencia:1994zi} which 
appear in the spin-spin correlation terms of the amplitude squared.
These terms involve the polarizations of {\it both} neutralinos, with
one polarization perpendicular to the production plane.
Such a normal polarization component is a genuine 
signal of CP violation (neglecting higher order effects due 
to final state interactions~\cite{Valencia:1994zi}).
The polarizations of the neutralinos can be analyzed in their decays, 
that's why we consider the leptonic channels  
\footnote{Note that generally  parity-conserving neutralino decays, like
    $\ti\chi^0_i \to Z \ti\chi^0_1$ 
    or $\ti\chi^0_i \to h \ti\chi^0_1$,
    would lead to vanishing  CP-sensitive observables.
    Due to the Majorana properties of the neutralinos,
    the left and right neutralino couplings to the $Z$ (and Higgs)
    have equal absolute values, and thus
    all spin- and spin-spin correlations would be lost, 
    see the discussion in Ref.~\cite{Dreiner:2007ay}.
}
\be{eq:decay}
\tilde\chi_i^0\to\tilde\ell^\pm_{L,R} + \ell^\mp~, \quad
\tilde\chi_j^0\to\tilde\ell_{L,R}^{\prime\pm}  + \ell^{\prime\mp}~,
\quad \ell,\ell^{\prime }=e,\mu~.
\ee
Due to angular momentum conservation, the decay distributions of 
the final leptons $\ell$, $\ell^{\prime}$ are correlated to each other, 
and thus allow us to probe the spin-spin correlations.

In a previous publication, we have analyzed in this way the CP-sensitive spin-spin
correlations for chargino production and decay~\cite{Bartl:2008fu}.
Other works done on CP-sensitive observables in neutralino pair production at the 
ILC have taken into account the decay of only one neutralino, where again the 
normal polarization component signals CP violation~\cite{neut,Kittel:2004rp}.
Even the potential of transverse beam polarizations for CP observables in
neutralino production has been analyzed~\cite{Bartl:2005uh,Trans}.
CP observables have also been studied in decays of neutralinos, which originate 
from sfermions~\cite{Bartl:2003ck}.

The paper is organized as follows.
In Section~\ref{Lagrangians}, we define the Lagrangians
and complex couplings for neutralino production.
In Section~\ref{cross section}, we present the analytical
formulae for the amplitude squared of neutralino 
production and decay.
In Section~\ref{identify},
we identify the T-odd products in the spin-spin terms of the amplitude squared. 
In Section~\ref{observables},
we define the CP-sensitive observables which probe these terms.  
We present numerical results in Section~\ref{numerics}, where we 
also estimate the measurability of the CP-sensitive observables.
We give a summary and the conclusions in Section~\ref{conclusion}.

\section{Lagrangians and complex couplings \label{Lagrangians}}

\begin{figure}[t]
\hspace{-2.cm}
\begin{minipage}[t]{6cm}
\begin{center}
{\setlength{\unitlength}{0.6cm}
\begin{picture}(5,5)
\put(-2.5,-8.5){\includegraphics{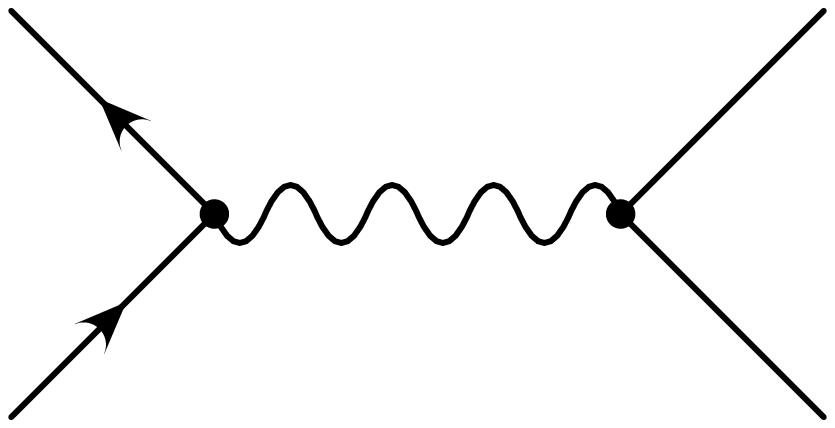}}
\put(1.7,-.4){{\small $e^{-}$}}
\put(7.8,-.4){{\small $\tilde\chi^0_j$}}
\put(1.7,3.8){{\small $e^{+}$}}
\put(7.8,3.8){{\small $\tilde\chi^0_i$}}
\put(5.4,2.5){{\small $Z$}}
\end{picture}}
\end{center}
\end{minipage}
\hspace{-0.5cm}
\begin{minipage}[t]{5cm}
\begin{center}
{\setlength{\unitlength}{0.6cm}
\begin{picture}(2.5,5)
\put(-4,-9){\includegraphics{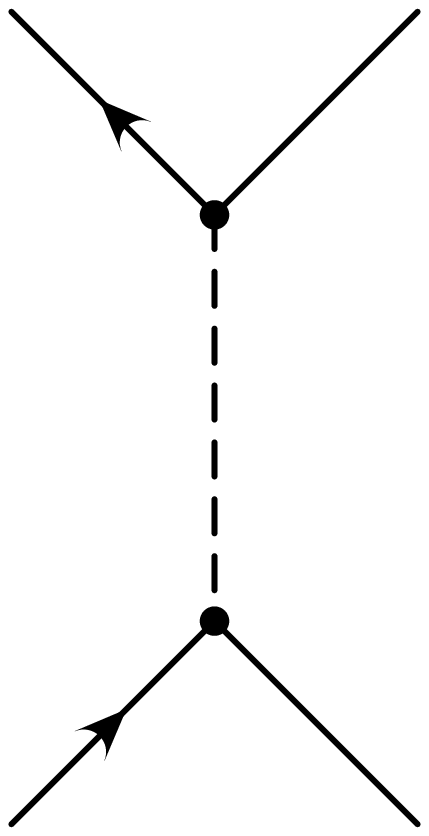}}
\put(1.8,-1.5){{\small $e^{-}$}}
\put(1.8,3.8){{\small $e^{+}$}}
\put(5.8,-1.5){{\small $\tilde\chi^0_j$}}
\put(5.8,3.8){{\small $\tilde\chi^0_i$}}
\put(4.4,1.5){{\small $\tilde{e}_{L,R}$}}
 \end{picture}}
\end{center}
\end{minipage}
\begin{minipage}[t]{5cm}
\begin{center}
{\setlength{\unitlength}{0.6cm}
\begin{picture}(2.5,5)
\put(-4.5,-9){\includegraphics{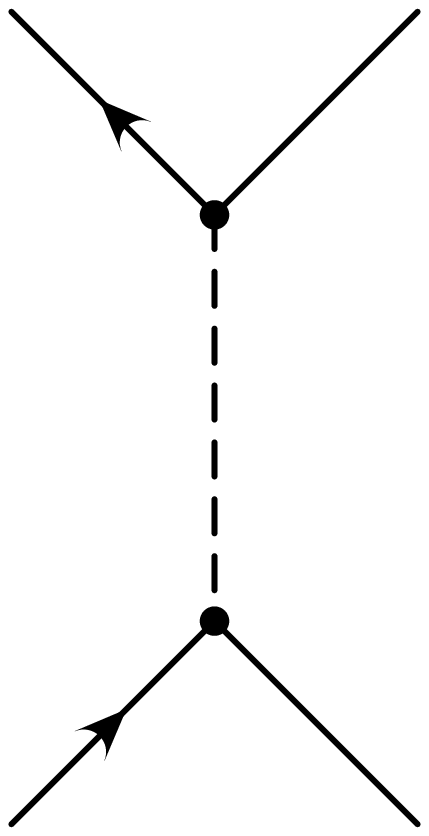}}
\put(1.3,-1.5){{\small $e^{-}$}}
\put(1.3,3.8){{\small $e^{+}$}}
\put(5.5,-1.5){{\small $\tilde\chi^0_i$}}
\put(5.5,3.8){{\small $\tilde\chi^0_j$}}
\put(3.9,1.5){{\small $\tilde{e}_{L,R}$}}
\end{picture}}
\end{center}
\end{minipage}
\vspace{.7cm}
\caption{\label{Fig:FeynProd}Feynman diagrams for neutralino production
  $e^{+}e^{-}\to\tilde{\chi}^0_i\tilde{\chi}^0_j$
\cite{Bartl:1986hp}.}
\end{figure}

In the MSSM, neutralino production 
$e^+e^-\to \tilde\chi^0_i\tilde\chi^0_j$
proceeds via  $Z$ boson exchange in the $s$-channel, and selectron 
$\tilde e_{L,R}$ exchange in the $t$- and $u$-channels, see the Feynman 
diagrams in Fig.~\ref{Fig:FeynProd}. 
The Lagrangians for production and decay 
are~\cite{Bartl:1986hp,Moortgat-Pick:1999di}
\begin{eqnarray}
{\cal L}_{Z e \bar e} &=&-\frac{g}{\cos\theta_W}
Z_{\mu}\bar e \gamma^{\mu}[L_e P_L+ R_e P_R]e ~,
\\[2mm]
{\cal L}_{Z\tilde{\chi}^0_i\tilde{\chi}^0_j} &=&
\frac{1}{2}\frac{g}{\cos\theta_W}
Z_{\mu}\bar{\tilde{\chi}}^0_i\gamma^{\mu}
[O_{ij}^{''L} P_L+O_{ij}^{''R} P_R]\tilde{\chi}^0_j~, \quad i, j=1,\dots,4~, 
\label{Zchichi}\\[2mm]
{\cal L}_{e \tilde{e}\tilde{\chi}^0_i} &=&
g f^L_{e i}\bar{e}P_R\tilde{\chi}^0_i\tilde{e}_L+
g f^R_{e i}\bar{e}P_L\tilde{\chi}^0_i\tilde{e}_R+\mbox{h.c.}~, 
\label{slechie}
\end{eqnarray}
with $P_{L, R}=(1\mp \gamma_5)/2$.
In the photino, zino, Higgsino basis
the couplings are~\cite{Moortgat-Pick:1999di} 
\begin{eqnarray}
        O_{ij}^{''L}&=&-\frac{1}{2} \left[
        (N_{i3}N_{j3}^*-N_{i4}N_{j4}^*)\cos2\beta
  +(N_{i3}N_{j4}^*+N_{i4}N_{j3}^*)\sin2\beta \right]~,\\[2mm]
O_{ij}^{''R}&=&-O_{ij}^{''L*}~,\\[2mm]
f_{e i}^L &=& \sqrt{2}\bigg[\frac{1}{\cos
\theta_W}(\frac{1}{2}-\sin^2\theta_W)N_{i2}+\sin \theta_W
N_{i1}\bigg]~,\\[2mm]
f_{e i}^R &=& \sqrt{2} \sin \theta_W
\Big[\tan \theta_W N_{i2}^*- N_{i1}^*\Big]~,\\[2mm]
 L_e&=&\sin^2\theta_W - \frac{1}{2}~, \quad
 R_e\;=\;\sin^2\theta_W \label{eq_5}~,
\label{eq_6}
\end{eqnarray}
with the weak mixing angle $\theta_W$,
the weak coupling constant $g=e/\sin\theta_W$, $e>0$,
and the ratio $\tan \beta=v_2/v_1$ of the vacuum expectation values 
of the two neutral Higgs fields.
The neutralino couplings $O_{ij}^{''L,R}$ and  $f^{L,R}_{e i}$ 
contain the complex mixing elements $N_{ij}$,
which diagonalize the neutralino matrix
 $N^*Y N^{\dagger}=
 {\rm diag}(m_{\chi_i^0})$~\cite{haberkane}, 
with the neutralino masses $ m_{\chi_i^0}>0$.
In the MSSM with CP violation, the couplings
$O_{ij}^{''L,R}$ and  $f^{L,R}_{e i}$ are in general complex due
to non-vanishing CP phases  $\phi_\mu$ and $\phi_{1}$. 
Here we adopt the standard convention that a possible phase
of $M_2$ can be absorbed by redefining the particle fields.

\section{Cross section \label{cross section}}

The differential cross section for neutralino production 
$e^+e^-\to\tilde\chi_i^0\tilde\chi_j^0$
and decay
$\tilde\chi_i^0\to\tilde\ell^\pm_{L,R} \ell^\mp$,
$\tilde\chi_j^0\to\tilde\ell_{L,R}^{\prime\pm} \ell^{\prime\mp}$,
can be written 
\be{eq:crossection}
{\rm d}\sigma=\frac{1}{2~s}~|T|^2~{\rm dLips}~,
\ee
with the center-of-mass energy $\sqrt s$, and the Lorentz invariant phase 
space element ${\rm dLips}$, see Appendix~\ref{Phase space}.
The amplitude squared $|T|^2$ was calculated in 
Ref.~\cite{Moortgat-Pick:1999di} in the spin density matrix
formalism\footnote{For a detailed discussion of the 
                   spin density matrix formalism, 
                   we refer to Ref.~\cite{Haber:1994pe}.}
\baq{eq:amplitudesquared}
|T|^2&=&4|\Delta(\tilde\chi_i^0)|^2|\Delta(\tilde\chi_j^0)|^2
\left[
P~D_i~D_j+
\sum^3_{a=1}\Sigma^a_P~\Sigma^a_{D_i}~D_j\right.\nonumber\\[2mm]
{}&&\left.+\sum^3_{b=1}\Sigma^b_P~\Sigma^b_{D_j}~D_i
+\sum^3_{a,b=1}\Sigma^{ab}_P~\Sigma^a_{D_i}~\Sigma^b_{D_j}
\right]~,
\eaq
with the neutralino propagators 
$\Delta(\tilde\chi_{i,j}^0)=1/[p^2_{\chi_{i,j}^0}-m^2_{\chi_{i,j}^0} 
+im_{\chi_{i,j}^0}\Gamma_{\chi_{i,j}^0}]$.
The amplitude squared has contributions
from neutralino production ($P$) and decay ($D$).
The terms $P$ and $D_i$, $D_j$ are those parts of the
spin density production and decay matrices, respectively,
that are independent of the polarizations of the neutralinos.
The contributions $\Sigma^a_P$ and $\Sigma^a_{D_i}$ depend on the polarization
basis vectors $s_{\chi_i}^a$ of neutralino $\tilde\chi^0_i$.
Similarly, $\Sigma^b_P$ and $\Sigma^b_{D_j}$ 
depend on the polarization basis vectors $s_{\chi_j}^b$ of 
neutralino $\tilde\chi^0_j$.
See Appendix~\ref{Vectors}, Eq.~\rf{eq:polvec}
for the explicit definition of the spin vectors.
We choose a coordinate frame such that $a,b =3$ denote the
longitudinal polarizations, $a,b =1$ the transversal
polarizations in the production plane, and $a,b =2$ the
polarizations normal to the production plane.
The decay terms $D_i$, $D_j$, $\Sigma^a_{D_i}$, and $\Sigma^b_{D_j}$ are given in
Appendix~\ref{Decaydensity}.
The full expressions for the production terms $P$, $\Sigma^a_P$, $\Sigma^b_P$, and
$\Sigma^{ab}_P$ can be found in Ref.~\cite{Moortgat-Pick:1999di}. 

\medskip

The contributions to the amplitude squared which depend on the polarizations 
of both neutralinos are the spin-spin correlation terms $\Sigma^{ab}_P$. 
The CP-sensitive parts of the spin-spin correlation terms include one neutralino 
spin vector with a component perpendicular to the production plane, i.e.,  
$ab=12,21,23,32$~\cite{Moortgat-Pick:1999di}
\begin{eqnarray}
\Sigma_P^{ab}(ZZ)&=&
             \frac{g^4}{\cos^4\theta_W}  |\Delta(Z)|^2 (R_e^2 c_R+L_e^2 c_L) 
		  ~{\rm Im}\{O^{''L}_{ij} O^{''R \ast}_{ij}\}~f^{ab}~,
        \label{eq:Toddpart1}\\[2mm]
\Sigma_P^{ab}(Z \tilde{e}_L) &=& 
	 \frac{g^4}{2\cos^2\theta_W} L_e c_L \Delta(Z) 
         [\Delta_u^{\ast}(\tilde{e}_L) +\Delta_t^{\ast}(\tilde{e}_L)]
        ~{\rm Im}\{f^{L}_{e i}f^{L\ast}_{e j} O^{''L}_{ij}\}~f^{ab}~,
      \label{eq:Toddpart2}\\[2mm]
\Sigma_P^{ab}(\tilde{e}_L \tilde{e}_L)&=& 
	  -\frac{g^4}{4} c_L  \Delta_{u}(\tilde{e}_L)\Delta_t^{\ast}(\tilde{e}_L)
        ~{\rm Im}\{(f^{L }_{e i})^2 (f^{L\ast }_{e j})^2 \}~f^{ab}~,
       \label{eq:Toddpart3}\\[2mm]
\Sigma_P^{ab}(Z \tilde{e}_R) &=& 
	 -\frac{g^4}{2\cos^2\theta_W} R_e c_R \Delta(Z) 
         [\Delta_u^{\ast}(\tilde{e}_R) +\Delta_t^{\ast}(\tilde{e}_R)]
        ~{\rm Im}\{f^{R}_{e i}f^{R\ast}_{e j} O^{''R}_{ij}\}~f^{ab}~,\qquad
      \label{eq:Toddpart4}\\[2mm]
\Sigma_P^{ab}(\tilde{e}_R \tilde{e}_R)&=& 
	  \frac{g^4}{4} c_R  \Delta_u(\tilde{e}_R)\Delta_t^{\ast}(\tilde{e}_R)
        ~{\rm Im}\{(f^{R }_{e i})^2 (f^{R\ast }_{e j})^2 \}~f^{ab}~.
       \label{eq:Toddpart5}
     \end{eqnarray}
The dependence on the longitudinal beam polarizations 
is given by 
\begin{equation}
c_L =(1-{\mathcal P}_-)(1+{\mathcal P}_+), \quad 
c_R= (1+{\mathcal P}_-)(1-{\mathcal P}_+)~,
\end{equation}
with ${\mathcal P}_-$ and ${\mathcal P}_+$ the degrees of 
longitudinal polarization of the electron and positron beam, 
respectively, with $-1\leq{\mathcal P}_\pm \leq 1$.
Generally the contributions from the exchange  
of $\tilde{e}_{R}$ ($\tilde{e}_{L}$) 
are enhanced and those of $\tilde{e}_{L}$ ($\tilde{e}_{R}$) are
suppressed for ${\mathcal P}_- > 0 , {\mathcal P}_+ < 0~
({\mathcal P}_-<0,{\mathcal P}_+>0)$.
%
The propagators are
$\Delta(Z)=i/(s-m^2_Z)$,
$\Delta_t(\tilde e_{L,R})=i/(t-m^2_{\tilde e_{L,R}})$, 
$\Delta_u(\tilde e_{L,R})=i/(u-m^2_{\tilde e_{L,R}})$,
with $s=(p_{e^-}+p_{e^+})^2$, $t=(p_{e^-}-p_{\chi_j})^2$, 
$u=(p_{e^-}-p_{\chi_i})^2$~\cite{Moortgat-Pick:1999di}.

The spin-spin correlation terms $\Sigma^{ab}_P$ in 
Eqs.~(\ref{eq:Toddpart1})--(\ref{eq:Toddpart5}) explicitly depend 
on the imaginary parts of the products of neutralino couplings,
${\rm Im}\{O^{''L}_{ij} O^{''R \ast}_{ij}\}$,
${\rm Im}\{f^{L}_{e i}f^{L\ast}_{e j} O^{''L}_{ij}\}$,
${\rm Im}\{(f^{L }_{e i})^2 (f^{L\ast }_{e j})^2 \}$,
${\rm Im}\{f^{R}_{e i}f^{R\ast}_{e j} O^{''R}_{ij}\}$, and
${\rm Im}\{(f^{R }_{e i})^2 (f^{R\ast }_{e j})^2 \}$.
For $i \neq j$ they are manifestly CP-sensitive, i.e., sensitive to the
phases $\phi_\mu$ and $\phi_{1}$ of the neutralino sector.
These imaginary parts of the couplings are multiplied by 
T-odd factors $f^{ab}$, which we discuss in detail 
in the next section.

\section{T-odd products of the spin-spin correlations \label{identify}}

The kinematical dependence of the spin-spin correlation terms
of neutralino production, Eqs.~(\ref{eq:Toddpart1})--(\ref{eq:Toddpart5}), 
is given by the T-odd function~\cite{Moortgat-Pick:1999di}
\baq{eq:f8}
f^{ab}&=&
\phantom{+}
(p_{e^+}\!\cdot\! p_{\chi_j})
[p_{e^-}, p_{\chi_i}, s_{\chi_i}^{a}, s_{\chi_j}^{b}]
+(p_{e^-} \!\cdot\!p_{\chi_i})
[p_{e^+}, p_{\chi_j}, s_{\chi_i}^{a}, s_{\chi_j}^{b}]
\nonumber\\
{}&&
+(p_{e^+}\!\cdot\! s_{\chi_j}^{b})
 [p_{e^-},  p_{\chi_i},  p_{\chi_j},  s_{\chi_i}^{a}]
+(p_{e^-} \!\cdot\!s_{\chi_i}^a)
[p_{e^+},  p_{\chi_i},  p_{\chi_j},  s_{\chi_i}^{b}]~,
\eaq
with the short hand notation of the 
epsilon product of the four four-vectors $p_i$  
\begin{equation}
 [p_{1},p_2,p_{3},p_{4}]  \equiv
\varepsilon_{\mu\nu\alpha\beta}~
      p_{1}^{\mu}~p_2^{\nu}~p_{3}^{\alpha}~p_{4}^{\beta}~,
\quad {\rm with} \quad \varepsilon_{0123}=-1.
\label{eq:epsilon}  
\end{equation}
Since each of the spacial components of the four-momenta 
or spins changes sign under a naive time transformation, $t \to -t$, 
the  epsilon product, and thus the function $f^{ab}$, is T-odd.
In Appendix~\ref{Vectors}, we give $f^{ab}$ also in the laboratory system.

In order to identify the T-odd products which appear in 
the spin-spin correlations of production and decay,
we analyze the corresponding terms of the amplitude squared, 
Eq.~\rf{eq:amplitudesquared}, in more detail 
\baq{eq:kinedep}
|T|^2 \; \supset \;
\sum^3_{a,b=1}\Sigma^{ab}_P~\Sigma^a_{D_i}~\Sigma^b_{D_j} \; \propto \; 
\sum^3_{a,b=1}f^{ab}\cdot ( p_{\ell}\!\cdot\! s^{a}_{\chi_i})
\cdot (p_{\ell'} \!\cdot\! s^{b}_{\chi_j} ) ~,
\eaq
where the scalar products $(p_{\ell}\cdot  s^{a}_{\chi_i})$ and  
$(p_{\ell'} \cdot  s^{b}_{\chi_j})$ stem from $\Sigma^a_{D_i}$ 
and $\Sigma^b_{D_j}$, respectively, see Eqs.~(\ref{neutdecay1}) or 
(\ref{neutdecay2}) in Appendix~\ref{Decaydensity}.  Using the explicit 
expression for $f^{ab}$, Eq.~\rf{eq:f8}, and the completeness relation for 
the neutralino spin vectors, Eq.~\rf{eq:chicompleteness}, the right-hand
side of of the second equation in Eq.~\rf{eq:kinedep} can be written as
\baq{eq:OT}
\Ta =
(p_{e^+} \!\cdot\! p_{\chi_j})
[ p_{e^-}, p_{\chi_i},  p_{\ell}, p_{\ell'}]~
+(p_{e^-}  \!\cdot\!p_{\chi_i})
 [p_{e^+},  p_{\chi_j}, p_{\ell}, p_{\ell'}]~~
\label{OT}
\nonumber\\[3mm]
+(p_{e^+}  \!\cdot\!p_{\ell'})
[ p_{e^-}, p_{\chi_i}, p_{\chi_j}, p_{\ell}]~
+(p_{e^-}  \!\cdot\!p_{\ell})
[p_{e^+}, p_{\chi_i}, p_{\chi_j}, p_{\ell'}]~.
\eaq
We have now identified the CP-sensitive terms  of the
neutralino spin-spin correlations.  
They are proportional to the T-odd product $\Ta$, Eq.~\rf{eq:OT},
which can now be used to define various CP asymmetries and
CP observables in neutralino production and decay.
Due to their similar kinematical dependence, the definition
of CP observables is analogous to those in 
chargino production and decay, see Ref.~\cite{Bartl:2008fu}.

\section{CP-sensitive observables \label{observables}}

In this Section, we define various CP-sensitive observables, which 
depend on the T-odd parts of the spin-spin correlations for neutralino 
production and decay. For an operator ${\mathcal  O}$, 
we define its expectation value by~\cite{Bartl:2008fu}
\be{eq:Obs}
\langle {\mathcal  O} \rangle=
\frac{ \int{\mathcal  O}~ |T|^2~ {\rm dLips} } 
     {\int |T|^2~ {\rm dLips} }
=
\frac{1}{\sigma} \int~{\mathcal  O}~ 
\frac{{\rm d}\sigma}{\rm dLips}~{\rm dLips}~.
\ee
If the operator  ${\mathcal O}$ is chosen of the form
like the T-odd terms $\Ta$, Eq.~\rf{eq:OT},
the CP-sensitive parts of the spin-spin correlations in 
neutralino production can be projected out
\baq{eq:Obsx}
\langle {\mathcal  O} \rangle=
\frac{ \int{\mathcal O}~ 
    \Sigma^{ab}_P~\Sigma^a_{D_i}~\Sigma^b_{D_j}~ {\rm dLips} } 
     {\int P~D_i~D_j~{\rm dLips} }~,
\eaq
with an implicit sum over $(a,b)= (1,2),(2,1),(2,3),(3,2)$.
In the numerator remain only the CP-sensitive parts of the spin-spin terms 
of the amplitude squared. Only they contain the T-odd product ${\mathcal O}$.
In the denominator, all spin- and spin-spin correlation terms vanish, and 
only the spin-independent part $P~D_i~D_j$ contributes. Note that for the 
phase space element $ {\rm dLips}$ in Eq.~\rf{eq:Obsx}, we have already 
used the narrow width approximation for the propagators, 
see Eq.~\rf{eq:narrowwidth}.

In general the largest observables are obtained by
using an operator ${\mathcal O}$, which exactly matches the kinematical 
dependence of the CP-sensitive terms in the amplitude squared, that is
${\mathcal O} =\Ta$, Eq.~\rf{eq:OT}.
In the literature, this technique is sometimes referred to 
\emph{optimal observables}~\cite{optimal}. 
Thus for the operator $\Ta$ we define the two CP-sensitive observables
\be{eq:Obs1}
\Oa  \qquad {\rm and} \qquad
\Aa =\langle {\rm Sgn}(\Ta)\rangle ~.
\ee
Neglecting higher order effects due to final state 
interactions~\cite{Valencia:1994zi}, the observable $\Aa$ is a CP asymmetry. 
It  is the expectation value for the sign of the T-odd product $\Ta$,
and can be written as 
\be{eq:ObsN}
\Aa = \frac{N_+-N_-}{ N_+  +N_-}~,
\ee
the difference of the number of events with positive ($N_+$) and negative ($N_-$) 
sign of $\Ta$, normalized by the total number of events  $N= N_+  +N_-$.
On the other hand, $\Oa$  is the expectation value of the momentum configuration 
$\Ta$ itself for the event sample.

Two further T-odd products were considered in Ref.~\cite{Bartl:2008fu}.
One product is obtained from $\Ta$, Eq.~\rf{eq:OT},
in replacing the four-momenta by the (normalized) three-momentum vectors
in the center-of-mass system, see  Appendix~\ref{Vectors},
\baq{eq:Todd1prod}
 \Tb=(\hat p_{e^-}\cdot \hat p_{\ell'}) ~
 \hat p_{e^-}\cdot(\hat p_{\chi_j}\times\hat p_{\ell})+
(\hat p_{e^-}\cdot \hat p_{\ell})~
 \hat p_{e^-}\cdot(\hat p_{\chi_j}\times\hat p_{\ell'})~,
\eaq
with $\hat p = \vec p/ |\vec p|$. In contrast to  $\Ta$, Eq.~\rf{eq:OT}, 
this product does not involve the energies of the neutralinos and
leptons. For the operator $\Tb$, we again define two CP-sensitive observables 
\be{eq:Obs2}
\Ob  \qquad {\rm and} \qquad
\Ab=\langle {\rm Sgn}(\Tb )\rangle ~.
\ee
Since both T-odd products $\Ta$ and $\Tb$ include the neutralino momentum 
$ p_{\chi_i}$ and/or $ p_{\chi_j}$, their experimental reconstruction is required. 
For the subsequent two-body decays of the neutralinos which we consider here,
the neutralino momentum three-vectors can be reconstructed up to a sign
ambiguity in their second component, if the masses of the involved particles 
are known, see for example Ref.~\cite{Bartl:2005uh}.

A T-odd product which does not depend on the neutralino momenta
is obtained by substituting on the right hand side of Eq.~\rf{eq:Todd1prod}
the neutralino three-momenta by the corresponding 
decay lepton three-momenta  $\vec{p}_{\chi_i}\to \vec{p}_{\ell}$ and
$\vec{p}_{\chi_j}\to \vec{p}_{\ell'}$~\cite{Bartl:2008fu},
\be{eq:Todd1}
\Tc =
\hat{p}_{e^-}\cdot(\hat{p}_{\ell}+\hat{p}_{\ell'})~
\hat{p}_{e^-}\cdot(\hat{p}_{\ell}\times \hat{p}_{\ell'})~.
\ee
Also for $\Tc$ we define a CP-sensitive observable and its corresponding asymmetry, 
\be{eq:Obs3}
\Oc \qquad {\rm and} \qquad
\Ac =\langle {\rm Sgn}(\Tc)\rangle ~.
\ee
Thus, depending on the type of correlation used, two classes of CP observables 
can be defined; one class requires the reconstruction of the neutralino momenta, 
the other class not. However, as we will show in the numerical section, the largest 
observables are obtained if indeed the neutralino momenta can be reconstructed.


\subsection{Relative signs of the CP observables \label{signdependence}}

Each of the above defined CP observables
depends in principle on the various decay channels of the two 
neutralinos. For each neutralino, these are
\begin{eqnarray}
 \tilde\chi^0_k &\to& \tilde\ell_R^+  + \ell^-,\\
                &\to& \tilde\ell_R^-  + \ell^+,
\end{eqnarray}
for $\ell=e,\mu$, and also
\begin{eqnarray}
 \tilde\chi^0_k &\to& \tilde\ell_L^+  + \ell^-,\\
                &\to& \tilde\ell_L^-  + \ell^+,
\end{eqnarray}
if the decay into the (usually) heavier left slepton is kinematically allowed.
However, only the sign of the CP observable 
changes, depending on the charge and the type (L or R) 
of the two decay sleptons, for an overview see Table~\ref{tableAsymm}.
The reason is that the signs of the corresponding two 
neutralino decay terms, $\Sigma^a_{D_i}$ and $\Sigma^b_{D_j}$,
only depend on the charge and the type of the two decay sleptons,
see Eqs.~(\ref{neutdecay1}) and (\ref{neutdecay2}), respectively.
The absolute value of an observable is independent 
of the particular decay channels, since the  
absolute values of the couplings $|f^{L,R}_{\ell i}|$ or $|f^{L,R}_{\ell j}|$
of the decay sleptons, as well as their masses,  
cancel in the numerator and denominator of Eq.~\rf{eq:Obsx}.
This means in turn that we have to distinguish from 
which neutralino $\tilde\chi^0_i$ or $\tilde\chi^0_j$
the final state leptons  $\ell$ and $\ell'$ originate.
Without that information, the contributions to the CP observables
from the final leptons with charge combinations $\ell^-\ell'^-$ and $\ell^+\ell'^+$
would identically cancel those contribution from 
 $\ell^+\ell'^-$ and $\ell^-\ell'^+$. 

Furthermore if also the decay into $\tilde\ell_L$ is kinematically possible, 
the type of the sleptons, $\tilde\ell_L$ or $\tilde\ell_R$, into which the neutralinos decay, 
has to be determined.
Such a discrimination can in principle be accomplished 
by using the different energy distributions of the decay leptons, 
since their kinematical limits depend on the mass difference of 
the decaying neutralino and slepton.

Note however, that if the final lepton momenta are assigned correctly,
one is able to reconstruct the production plane. 
Although the two lightest neutralinos in the end of the decay chains,
$\tilde\chi^0_k \to \tilde\ell \ell$,
$\tilde\ell\to \tilde\chi^0_1  \ell$,
carry away their missing momentum, the ambiguities in the azimuthal angles
of the produced neutralinos can be resolved with a measurement
and correct assignment of the four final lepton momenta,
see the discussion in Ref.~\cite{Bartl:2005uh}.
Certainly the feasibility of such an event reconstruction
can only be answered by a detailed experimental analysis, 
which is however beyond the scope of the present work.

\begin{table}[t]
\caption{Relative signs of the CP-sensitive 
         observables for different decay combinations of neutralino 
         $\tilde\chi_i^0\to \tilde\ell_{L(R)}^\pm \ell^\mp$ (top row), 
         and neutralino $\tilde\chi_j^0\to \tilde\ell_{L(R)}^\pm \ell^{'\mp}$ 
         (left column).
\label{tableAsymm}}
\begin{center}
\begin{tabular}{c|c|c|c|c} 
 & $\tilde\ell_R^+$
 & $\tilde\ell_R^-$
 & $\tilde\ell_L^+$
 & $\tilde\ell_L^-$
\\\hline
$\tilde\ell_R^+$
& $+$ & $-$ & $-$ & $+$  
\\\hline
$\tilde\ell_R^-$
& $-$ & $+$ & $+$ & $-$
\\\hline
$\tilde\ell_L^+$
& $-$ & $+$ & $+$ & $-$
\\\hline
$\tilde\ell_L^+$
& $+$ & $-$ & $-$ & $+$
\end{tabular}
\\[5.0ex]
\end{center}
\end{table}

\subsection{Theoretical statistical significances\label{statistics}}

We have defined various kinds of CP-sensitive observables, which are based 
on the different T-odd products ${\mathcal O}=\Ta, \Tb, \Tc$. They either 
include $(\Ta,\Tb)$ or not include $(\Tc)$ the neutralino momentum.
In order to be able to compare these observables quantitatively,
we define their theoretical statistical significances.
A comparison of the numerical values of $\langle {\mathcal O}\rangle$ and 
${\mathcal A}=\langle {\rm Sgn}({\mathcal O})\rangle$ alone cannot be used
to decide which observable is more sensitive to the CP phases.
In addition, we are sometimes facing situations where large CP observables 
and asymmetries correspond to processes with small neutralino production
cross sections or branching ratios, and vice versa. Such effects can be considered
by combining both the CP observable and the cross section into one 
statistical quantity.

We define the theoretical statistical significance of the CP observable 
$\langle \bar {\mathcal { O}}\rangle$, where
$\bar {\mathcal O}={\mathcal O}$, or
$\bar {\mathcal  O}={\rm Sgn}({\mathcal O})$,
by~\cite{Bartl:2006bn,Bartl:2005uh}
\be{eq:EffAsy}
 S[\bar{\mathcal  O}]=\sqrt{N}~
\frac{|\langle \bar{\mathcal O}\rangle|}
     {\sqrt{\langle \bar{\mathcal O}^2\rangle}}~.
\ee
For neutralino production and decay the number of events  is
\baq{eq:Eventnumber}
N=F_N\times {\mathcal L}\times\sigma(e^+e^-\to \tilde\chi^0_i\tilde\chi^0_j)
\!\!\!&\times& \!\!\!\left[{\rm BR}(\tilde\chi^0_i\to \tilde e_R^+~e^-)
\times {\rm BR}(\tilde\chi^0_j\to \tilde e^{+}_R~ e^{-})\nonumber \right.\\[2mm]
&&{}\left. \!\!\!\!\!\!+
{\rm BR}(\tilde\chi^0_i\to \tilde e_L^+~e^-)
\times {\rm BR}(\tilde\chi^0_j\to \tilde e^{+}_L~e^{-})\nonumber\right.\\[2mm]
&&{}\left. \!\!\!\!\!\!+
{\rm BR}(\tilde\chi^0_i\to \tilde e_R^+~e^-)
\times {\rm BR}(\tilde\chi^0_j\to \tilde e^{+}_L~e^{-})\nonumber\right.\\[2mm]
&&{}\left. \!\!\!\!\!\!+
{\rm BR}(\tilde\chi^0_i\to \tilde e_L^+~e^-)
\times {\rm BR}(\tilde\chi^0_j\to \tilde e^{+}_R~e^{-})\right],
\eaq
with the integrated luminosity ${\mathcal L}$.
The combinatorial factor $F_N$ takes into account all possible neutralino decays
into sleptons with different flavors and charges.
We assume that the branching ratios of the neutralinos do not depend on them, i.e.,
${\rm BR}(\tilde\chi^0_k\to \tilde e^{+}_{n}~e^{-})=
 {\rm BR}(\tilde\chi^0_k\to \tilde e^{-}_{n}~e^{+})=
 {\rm BR}(\tilde\chi^0_k\to \tilde \mu^{+}_{n}~\mu^{-})=
 {\rm BR}(\tilde\chi^0_k\to \tilde \mu^{-}_{n}~\mu^{+})$,
 for $n=L$ and $R$. 
The  combinatorial factor is thus $F_N=4\times4=16$,
if we sum the two lepton flavors $e, \mu$ and the two charges,
$ \tilde\ell^{+}_{n}$ and $ \tilde\ell^{-}_{n}$.

The statistical significance  $S$ is equal to the number of 
standard deviations to which the corresponding CP observable
can be determined to be non-zero over statistical fluctuations. 
For example, $S = 1$ implies a measurement at the statistical 68\% confidence level, 
assuming an ideal detector, and full reconstruction of signal 
and background. Thus our definition of $S$ is theoretically motivated, 
and can only be regarded as as an upper bound on the confidence level
which at best can be obtained.
In order to give realistic values of the statistical significances, 
a detailed experimental study would be required, which is however
beyond the scope of the present work.

Also higher order corrections have to be included in a
comprehensive analysis. Although we expect the influence of electroweak corrections 
to our observables and  asymmetries to be small, the corrections to the neutralino masses and 
production cross sections can be $10\%$ at one-loop level~\cite{Oller:2005xg}.
The neutralino branching ratios for two-body decays may receive  
CP-even one-loop corrections of up to $16\%$ in some cases~\cite{Drees:2006um}. 
For chargino production additional CP-sensitive terms contribute at higher order
to the production cross section, which has recently been discussed  
in Ref.~\cite{Osland:2007xw}.

\section{Numerical results \label{numerics}}

We present numerical results for the CP-sensitive observables and asymmetries
for neutralino production  $e^+e^-\to\tilde\chi_2^0\tilde\chi_3^0$, and decay
$\tilde\chi_2^0\to\tilde\ell^\pm_{R} \ell^\mp$,
$\tilde\chi_3^0\to\tilde\ell_{R}^{\prime\pm} \ell^{\prime\mp}$, 
for $\ell,\ell^\prime=e,\mu$.
We will study the dependence of the CP observables on the phases and moduli of the higgsino 
and U(1) gaugino mass parameters $\mu=|\mu|e^{i\phi_{\mu}}$ and $M_1=|M_1|e^{i\phi_1}$,
respectively, in the framework of the general MSSM. In this model the restrictions on the
phases from the electron and neutron EDMs are less severe compared to the 
constrained MSSM~\cite{EDM}. Thus we will not take the EDMs into account, 
and show the full phase dependence of the observables.

The results are calculated with a center-of-mass energy of $\sqrt{s}=500$~GeV.
We choose longitudinal beam polarizations 
$\left(\mathcal P_-,\mathcal P_+\right)=\left(0.9,-0.6\right)$,
which enhance the $\tilde{e}_R$ exchange contribution.
The feasibility of measuring the observables also depends on the neutralino 
production cross section and decay branching ratios, which we discuss in
detail. For a comparison of the CP observables, and for giving an upper bound on the
confidence levels, we also present a theoretical estimate of their statistical significances.

For the neu\-tra\-lino widths and branching ratios, 
we include the two-body decays~\cite{Kittel:2004rp}
\begin{eqnarray}
\tilde\chi^0_i &\to& 
\tilde\ell_n   + \ell,~\quad
\tilde\nu_\ell + \nu_\ell,~\quad
 Z+\tilde\chi^0_m,~\quad
 h+\tilde\chi^0_m,~\quad
 W^\pm +\tilde\chi^\mp_k,
\label{neutdecaymodes}
\end{eqnarray}
with $m<i$; $k=1,2$;  $n=R,L$ for $\ell=e,\mu$, and $n=1,2$ for $\ell=\tau$.
We neglect three-body decays. We use the GUT inspired relation 
$|M_1|=5/3 M_2\tan^2\theta_W$, such that the dependence of the CP observables on the 
modulus of $M_1$ is investigated by using $M_2$. 
In order to reduce the number of free parameters further,  
we parametrize the slepton masses by $M_2$, and $m_0=80$~GeV fixed, which enter 
in the approximate solutions to the renormalization group equations, 
see Appendix~\ref{Decaydensity}.
We take stau mixing into account, and fix the trilinear scalar coupling 
parameter $A_\tau=250$~GeV. We use the SM parameters
$\sin^2\theta_W=0.2315$, $m_W =  80.41~{\rm GeV}$,
$m_Z =  91.187~{\rm GeV}$, $\alpha = 7.8125 \times 10^{-3}$.

The CP-sensitive neutralino coupling factors in
Eqs.~(\ref{eq:Toddpart1})--(\ref{eq:Toddpart5}) are zero for $i= j$, or vanishing phases 
$\phi_\mu$ and $\phi_1$. They are largest for $\phi_1=0.5\pi$ (or $1.5\pi$),  and for
a strong gaugino-higgsino mixing $M_2\approx|\mu|$. We find that a small
value of $\tan\beta$ is preferred to have large
CP observables and large significances. Therefore we center our numerical
discussion around a scenario with $\tan\beta=3$ and a strong gaugino-higgsino mixing.
The parameters are summarized in Tab.~\ref{tab:scenario}.
The corresponding particle masses, branching ratios,
and the cross section  are listed in Tab.~\ref{tab:masses}.
For this scenario, we analyze the phase dependence of the CP observables,
and then their dependence on $|\mu|$ and $M_2$.

\begin{table}[t]
\caption{
      Benchmark scenario for $e^+e^-\to\tilde\chi_2^0\tilde\chi_3^0$, 
      and decay $\tilde\chi_2^0\to\tilde\ell^\pm_{R} \ell^\mp$,
      $\tilde\chi_3^0\to\tilde\ell_{R}^{\prime\pm} \ell^{\prime\mp}$, 
      for $\ell,\ell^\prime=e,\mu$, at $\sqrt{s}=500$~GeV with
      beam polarizations 
    $\left(\mathcal P_-,\mathcal P_+\right)=\left(0.9,-0.6\right)$.
 \label{tab:scenario}}
\begin{center}
\begin{tabular}{cccccc} \hline
 
  \multicolumn{1}{c}{$M_2$} 
& \multicolumn{1}{c}{$|\mu|$}
& \multicolumn{1}{c}{$\phi_\mu$}  
& \multicolumn{1}{c}{$\phi_{1}$}  
& \multicolumn{1}{c}{$\tan{\beta} $}
& \multicolumn{1}{c}{$m_0$}
\\\hline
 
  \multicolumn{1}{c}{$270~{\rm GeV}$} 
& \multicolumn{1}{c}{$200~{\rm GeV}$}
& \multicolumn{1}{c}{$0$}  
& \multicolumn{1}{c}{$0.5\pi$} 
& \multicolumn{1}{c}{$3$}
& \multicolumn{1}{c}{$80~{\rm GeV}$}
\\\hline
\end{tabular}
\end{center}
\end{table}

\begin{table}[t]
\renewcommand{\arraystretch}{1.2}
\caption{SUSY masses, neutralino branching ratios and production cross section,
         for the benchmark scenario. The branching ratios are summed over  
         $\ell=e,\mu$ and both slepton charges.
         \label{tab:masses}}
\begin{center}
      \begin{tabular}{|c|c|c|}
\hline
$   m_{\chi^0_1}  =   121~{\rm GeV}$ &
$ m_{\tilde{e}_R} =   157~{\rm GeV}$ & 
${\rm BR}(\tilde\chi_2^0\to \tilde \ell_R \ell) = 66 \%$ \\
\hline
$   m_{\chi^0_2}  =   171~{\rm GeV}$ &
$ m_{\tilde{e}_L} =   256~{\rm GeV}$ & 
${\rm BR}(\tilde\chi_2^0\to \tilde \tau_1 \tau) = 34 \%$ \\
\hline
$   m_{\chi^0_3}  =   207~{\rm GeV}$ &
$m_{\tilde\tau_1} =   157~{\rm GeV}$ & 
${\rm BR}(\tilde\chi_3^0\to \tilde \ell_R \ell) = 66 \%$ \\
\hline
$ m_{\chi^\pm_1}  =   158~{\rm GeV}$ &
$m_{\tilde\tau_2} =   256~{\rm GeV}$ & 
${\rm BR}(\tilde\chi_3^0\to \tilde \tau_1 \tau) = 34 \%$ \\
\hline
$ m_{\chi^\pm_2}  =   318~{\rm GeV}$ &
$ m_{\tilde\nu}   =   246~{\rm GeV}$ & 
$  \sigma(e^+e^-\to\tilde\chi_2^0\tilde\chi_3^0) = 79~{\rm fb} $\\
\hline
\end{tabular}
\end{center}
\renewcommand{\arraystretch}{1.0}
\end{table}

\subsection{Phase dependence \label{phasedependence}}

\begin{figure}[t]
  \centering
  \subfigure[]{%
    \label{fig:AsyComp_phiM1}%
    \includegraphics[width=0.45\textwidth]{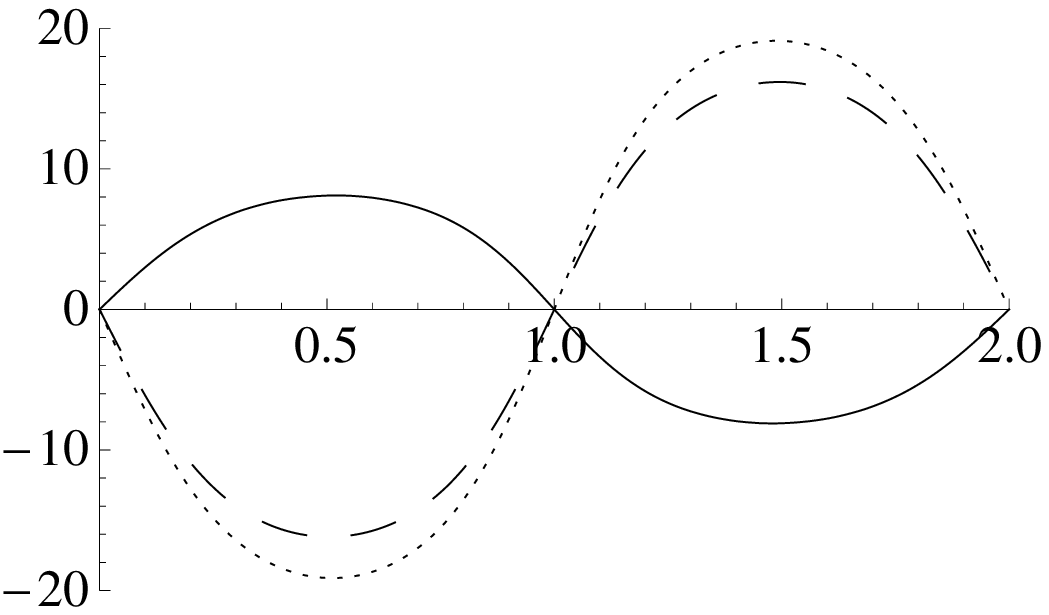}
    \put(0,40){$\phi_1[\pi]$}
    \put(-205,130){${\mathcal A}[\%]$}
    \put(-55,115){$\Aa$}
    \put(-55,85){$\Ab$}
    \put(-145,85){$\Ac$}
  }
  \hspace{10mm}
  \subfigure[]{%
    \label{fig:AsyComp_phiMu}%
    \includegraphics[width=0.45\textwidth]{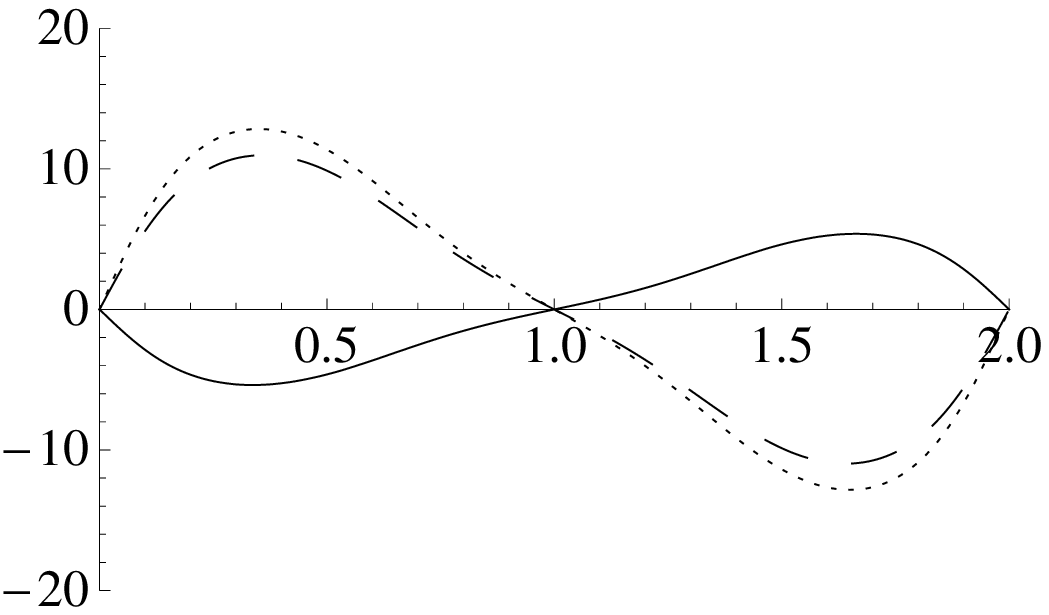}
    \put(0,40){$\phi_\mu[\pi]$}
    \put(-205,130){${\mathcal A}[\%]$}
    \put(-145,100){$\Aa$}
    \put(-155,70){$\Ab$}
    \put(-55,80){$\Ac$}
  }
\caption{\label{fig:AsyComp_phiM1Mu}
      Dependence of the CP asymmetries $\Aa$~(dotted), $\Ab$~(dashed),
      and $\Ac$~(solid), (a) on the phase $\phi_1$ with $\phi_\mu=0$,
      and (b) on the phase  $\phi_\mu$ with $\phi_1=\pi$, and the other
      parameters as defined in Tab.~\ref{tab:scenario}.
      }
\end{figure}

\begin{figure}[htbp]
  \centering
  \subfigure[]{%
    \label{fig:Asy3_phiphi}%
    \includegraphics[width=0.4\textwidth]{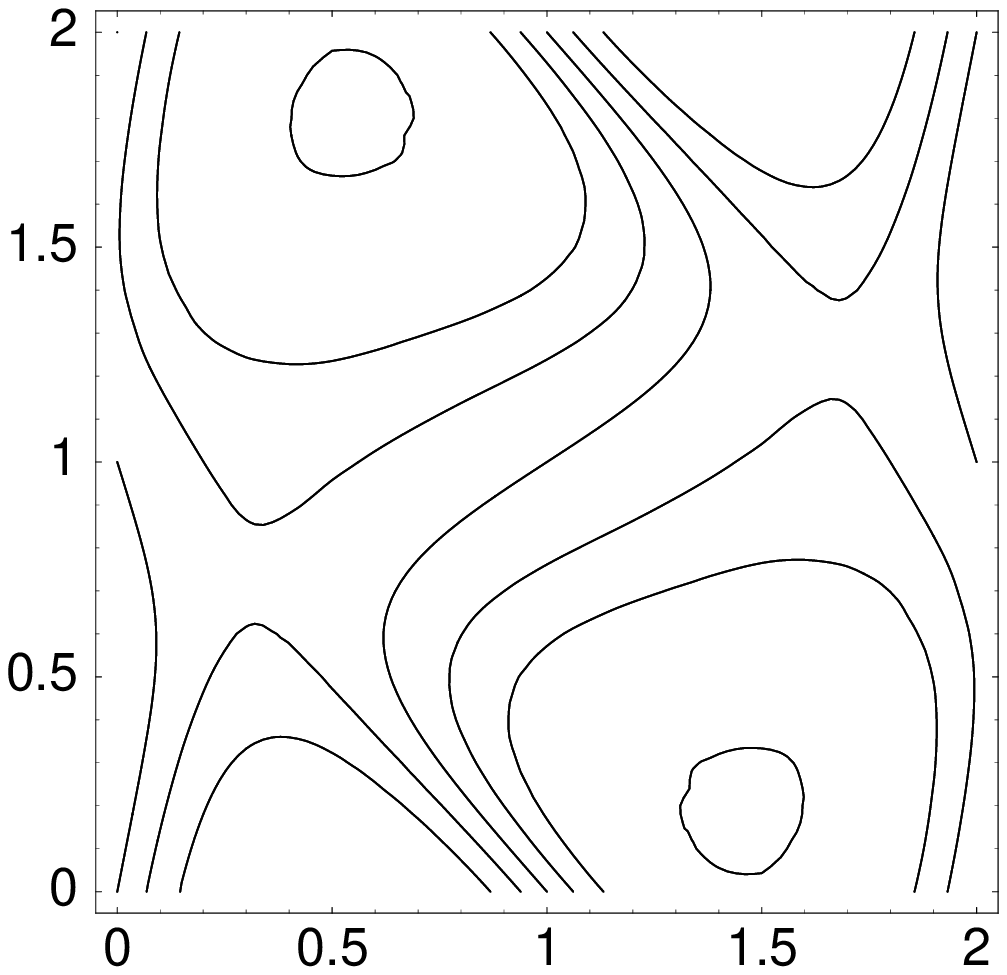}
    \put(-100,190){\fbox{$\Aa$ in \%}}
    \put(-25,-15){$\phi_1[\pi]$}
    \put(-200,165){$\phi_\mu[\pi]$}
    \put(-120,135){\footnotesize$-20$}
    \put(-105,122){\footnotesize$-10$}
    \put(-100,95){\footnotesize$-5$}
    \put(-75,85){\footnotesize$0$}
    \put(-50,95){\footnotesize$5$}
    \put(-30,60){\footnotesize$10$}
    \put(-50,25){\footnotesize$20$}
    \put(-155,58){\footnotesize$0$}
    \put(-130,65){\footnotesize$-5$}
    \put(-135,30){\footnotesize$-10$}
    \put(-10,123){\footnotesize$0$}
    \put(-28,129){\footnotesize$5$}
    \put(-40,145){\footnotesize$10$}
  }
  \hspace{15mm}
  \subfigure[]{%
    \label{fig:Sig3_phiphi}%
    \includegraphics[width=0.4\textwidth]{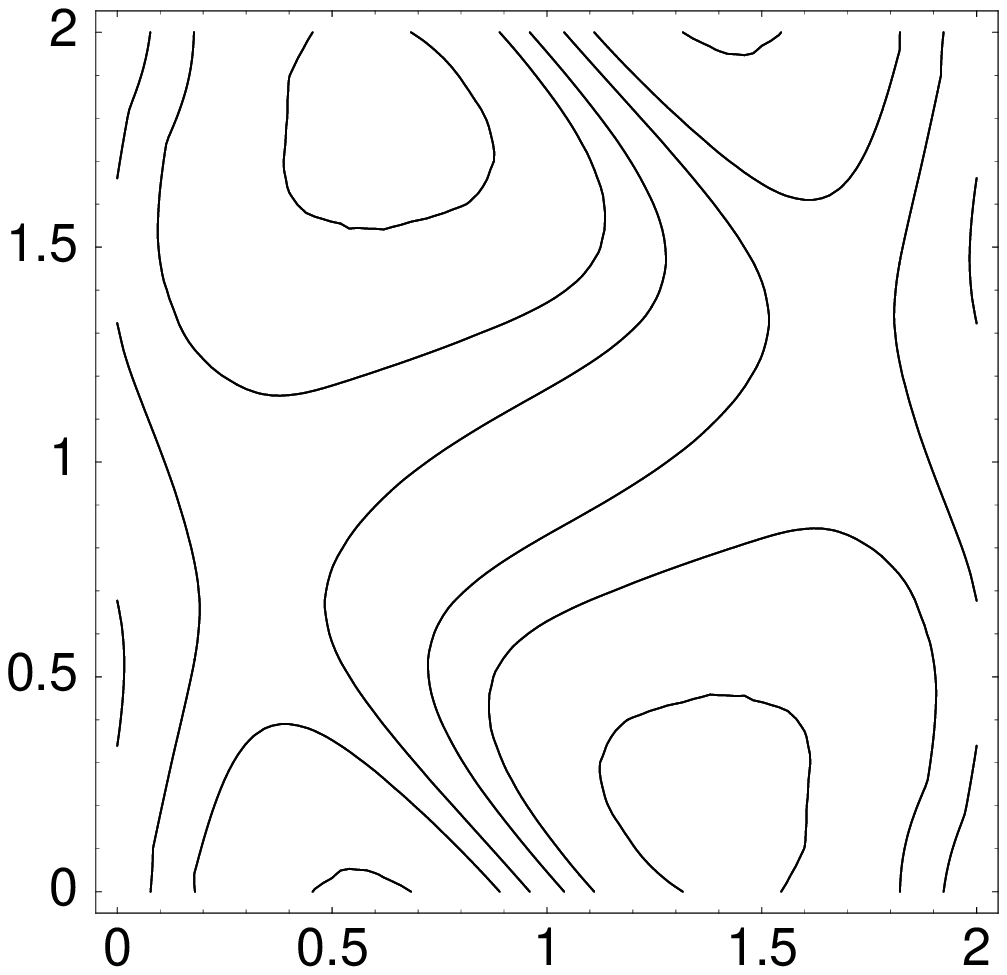}
    \put(-100,190){\fbox{$\SAa $}}
    \put(-25,-15){$\phi_1[\pi]$}
    \put(-200,165){$\phi_\mu[\pi]$}
    \put(-120,140){\footnotesize$25$}
    \put(-105,115){\footnotesize$13$}
    \put(-105,95){\footnotesize$5$}
    \put(-155,70){\footnotesize$5$}
    \put(-140,80){\footnotesize$5$}
    \put(-130,50){\footnotesize$13$}
    \put(-118,25){\footnotesize$25$}
    \put(-42,158){\footnotesize$25$}
    \put(-38,143){\footnotesize$13$}
    \put(-65,93){\footnotesize$5$}
    \put(-25,100){\footnotesize$5$}
    \put(-10,110){\footnotesize$5$}
    \put(-45,70){\footnotesize$13$}
    \put(-55,40){\footnotesize$25$}
  }\\ \vspace{15mm}
  \subfigure[]{%
    \label{fig:Asy1_phiphi}%
    \includegraphics[width=0.4\textwidth]{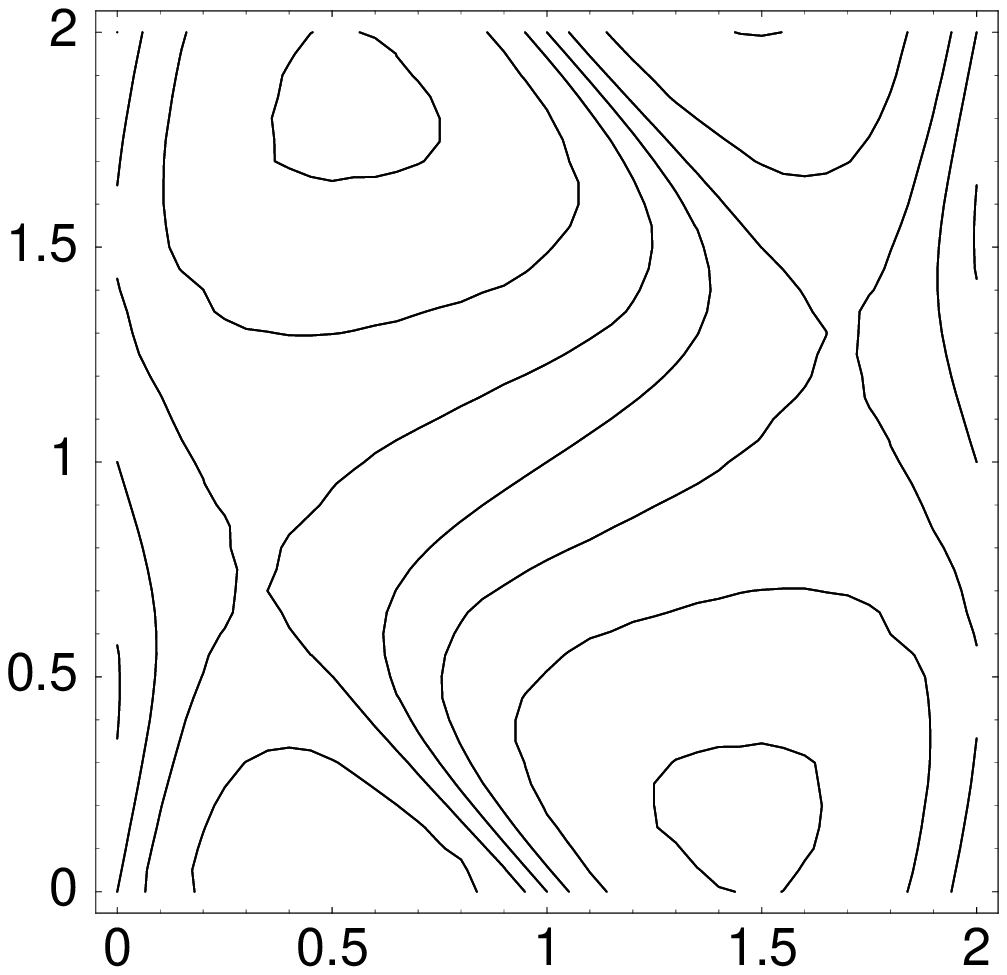}
    \put(-100,190){\fbox{$ \Ac $ in \%}}
    \put(-25,-15){$\phi_1[\pi]$}
    \put(-200,165){$\phi_\mu[\pi]$}
    \put(-125,150){\footnotesize$9$}
    \put(-105,125){\footnotesize$5$}
    \put(-105,103){\footnotesize$2$}
    \put(-75,85){\footnotesize$0$}
    \put(-55,100){\footnotesize$-2$}
    \put(-40,60){\footnotesize$-5$}
    \put(-50,25){\footnotesize$-9$}
    \put(-150,80){\footnotesize$0$}
    \put(-140,94){\footnotesize$2$}
    \put(-125,32){\footnotesize$5$}
    \put(-40,150){\footnotesize$-5$}
    \put(-30,85){\footnotesize$-2$}
    \put(-8,110){\footnotesize$0$}
  }
  \hspace{15mm}
  \subfigure[]{%
    \label{fig:Sig1_phiphi}%
    \includegraphics[width=0.4\textwidth]{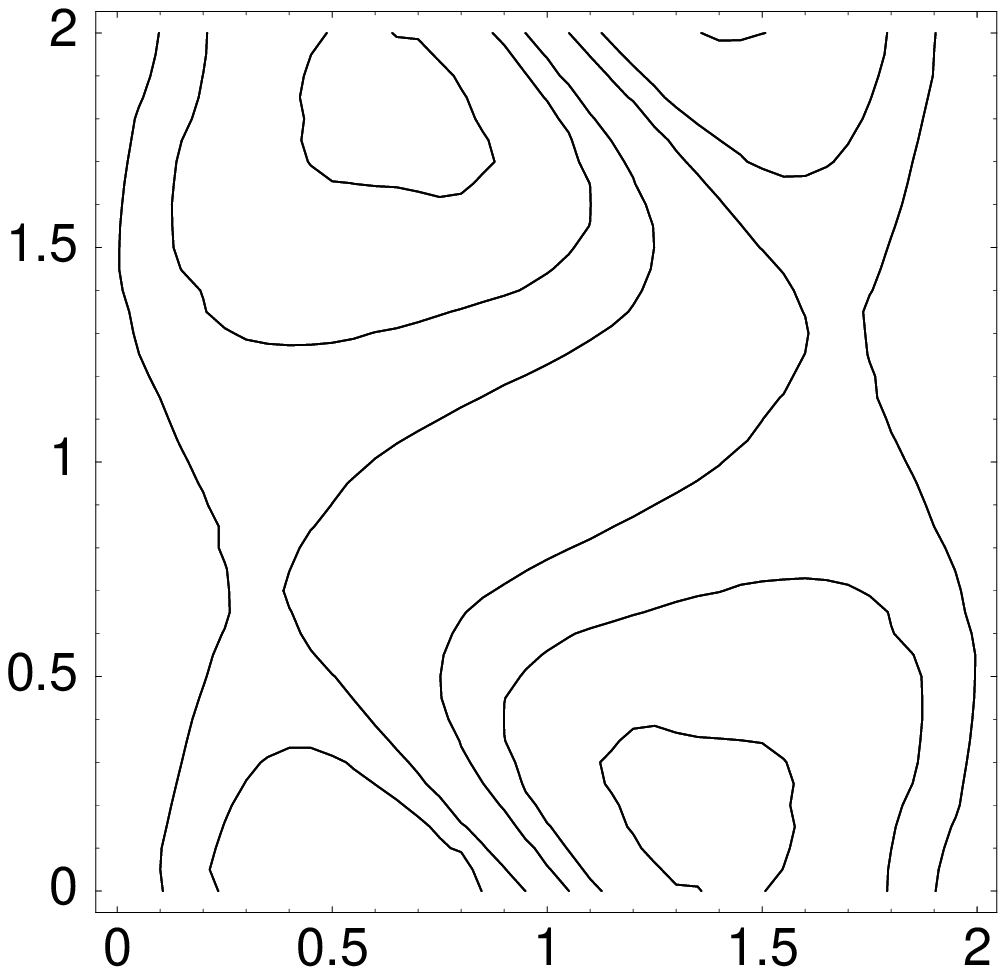}
    \put(-100,190){\fbox{$ \SAc $}}
    \put(-25,-15){$\phi_1[\pi]$}
    \put(-200,165){$\phi_\mu[\pi]$}
    \put(-120,150){\footnotesize$12$}
    \put(-98,123){\footnotesize$7$}
    \put(-105,100){\footnotesize$3$}
    \put(-148,83){\footnotesize$3$}
    \put(-125,32){\footnotesize$7$}
    \put(-38,146){\footnotesize$7$}
    \put(-48,100){\footnotesize$3$}
    \put(-20,110){\footnotesize$3$}
    \put(-35,76){\footnotesize$7$}
    \put(-55,35){\footnotesize$12$}
  }
\caption{\label{fig:AsySig_phiphi}
       Contour lines of the asymmetries $\Aa$ and $\Ac$, and
       their statistical significances in the $\phi_1$--$\phi_\mu$ plane,
       with an integrated luminosity of $\mathcal L=500$ fb$^{-1}$,
       for the scenario as defined in Tab.~\ref{tab:scenario}.
       }
\end{figure}

In Fig.~\ref{fig:AsyComp_phiM1}, we show the $\phi_1$ dependence of the CP 
asymmetries $\Aa$~\rf{eq:Obs1}, $\Ab$~\rf{eq:Obs2}, and $\Ac$~\rf{eq:Obs3}.
The asymmetries vanish at $\phi_1=0,\pi, 2\pi$, where the neutralino couplings are real.
They obtain largest values at $\phi_1\approx 0.5\pi$ and $\phi_1\approx 1.5\pi$
of about $\Aa=\pm 19\%$, $\Ab=\pm 16\%$, and $\Ac =\pm 8\%$.
In Fig.~\ref{fig:AsyComp_phiMu}, we  show the same asymmetries as a function of 
$\phi_\mu$, setting  $\phi_1=\pi$. They show a similar behavior, and again $\Aa$
attains the largest values of all three asymmetries. 
We do not present plots of the corresponding observables, since they show similar 
phase dependences as their corresponding asymmetries. They obtain values of 
$\Oa=-2.56 \times 10^{11}~{\rm GeV}^6$, $\Ob=-0.062$, and $\Oc=0.027$, for the 
scenario defined in Tab.~\ref{tab:scenario}.
\begin{figure}[htbp]
  \centering
  \subfigure[]{%
    \label{fig:SigComp1_phiM1}%
    \includegraphics[width=0.45\textwidth]{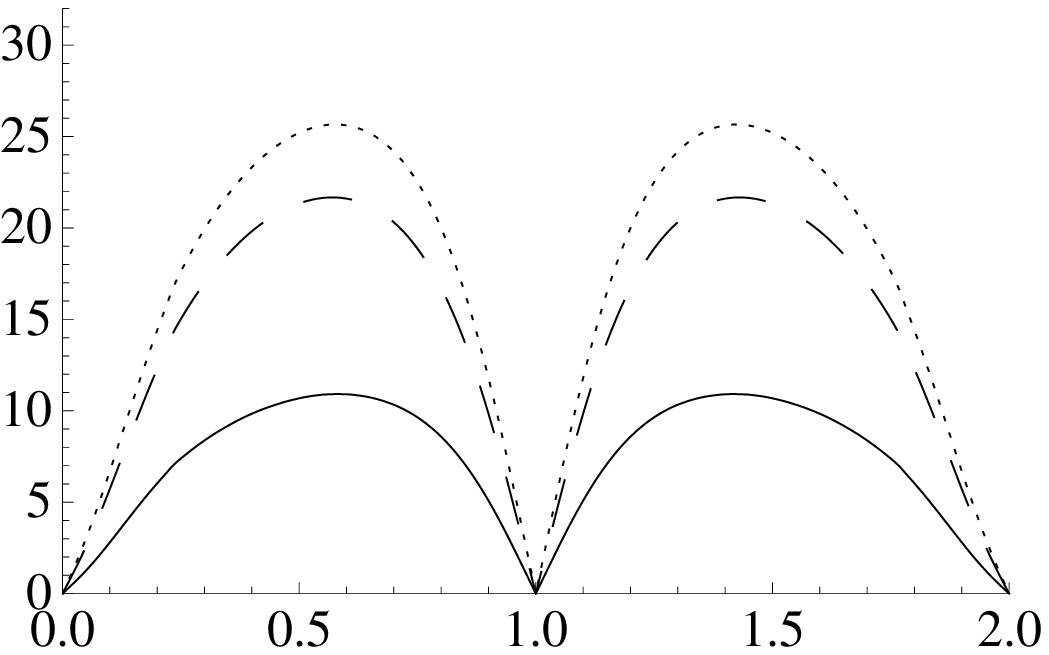}
    \put(0,-5){$\phi_1[\pi]$}
    \put(-210,130){$S[{\mathcal A}]$}
    \put(-150,110){$\SAa$}
    \put(-150,65){$\SAb$}
    \put(-150,30){$\SAc$}
  }
  \hspace{10mm}
  \subfigure[]{%
    \label{fig:SigComp2_phiM1}%
    \includegraphics[width=0.45\textwidth]{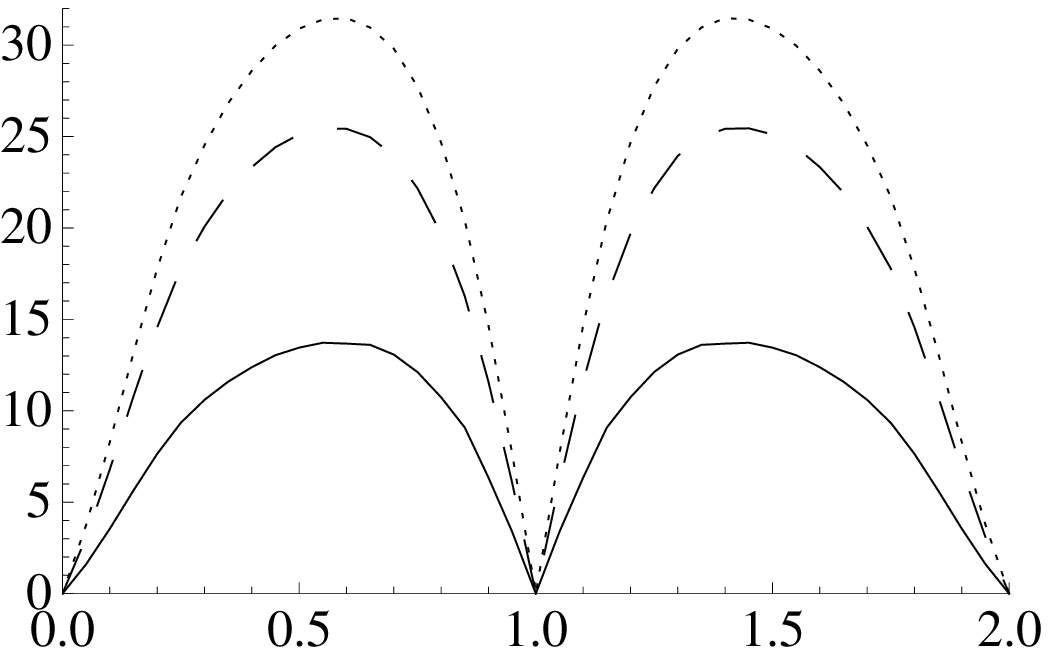}
    \put(0,-5){$\phi_1[\pi]$}
    \put(-210,130){$S[\mathcal O]$}
    \put(-150,130){$\SOa$}
    \put(-150,77){$\SOb$}
    \put(-150,35){$\SOc$}
  }
  \caption{\label{fig:SigComp_phiM1}
      Phase-dependence of the significances
      of (a) the asymmetries $\Aa$~(dotted), $\Ab$~(dashed), $\Ac$~(solid),
      and (b) of the observables $\Oa$~(dotted), $\Ob$~(dashed), $\Oc$~(solid),
      with an integrated luminosity of $\mathcal L=500$ fb$^{-1}$, 
      for the scenario as defined in Tab.~\ref{tab:scenario}.
      }
\end{figure}
In order to compare now the CP asymmetries ${\mathcal A}$ with their 
corresponding  CP observables $\langle{\mathcal O}\rangle$, we show their theoretical
significances $S$ as a function of $\phi_1$ in Fig.~\ref{fig:SigComp_phiM1}.
First we observe that the observables, Fig.~\ref{fig:SigComp2_phiM1},  
have  generally larger significances than their counterpart asymmetries, 
Fig.~\ref{fig:SigComp1_phiM1}. Secondly, the observables and asymmetries 
which are based on the T-odd products $\Ta$ and $\Tb$, which include the neutralino 
momentum, have the largest significances. They would be best suited for measuring 
CP phases  in the neutralino spin-spin correlations. The significance of $\Oa$ 
is twice as large as that of $\Oc$.
However, for their measurement a reconstruction of the neutralino momenta,
i.e., the production plane is necessary, which will be experimentally more 
involved. The need to only reconstruct the final state leptons might be an 
advantage in a realistic experimental environment. 
However, a detailed investigation which observable will be best suited
can only be answered by a thorough experimental analysis, which is beyond 
the scope of the present work.
In order to further illustrate the different magnitudes of the asymmetries $\Aa$ and $\Ac$,
we show them and the corresponding significances
as a function of the phases $\phi_\mu$ and $\phi_1$ in Fig.~\ref{fig:AsySig_phiphi}.

Finally, it should be noted that a measurement of observables which depend 
only on $\phi_\mu$ will be helpful to disentangle CP-violating effects in the 
neutralino system, which could originate  both from $\phi_1$ and $\phi_\mu$. 
This could be possible by investigating CP observables in the chargino 
system~\cite{Bartl:2008fu,CHAR} which solely depend on $\phi_\mu$.
Finally a global fit of CP-even~\cite{Moortgat-Pick:1999di,Kneur:1999nx} and 
CP-odd~\cite{neut,Kittel:2004rp,Bartl:2005uh,Trans} observables in the 
neutralino system could allow for a complete determination of the phases.

%
\subsection{ $\mu$ and $M_2$ dependence \label{mum1dependence}}

In order to estimate the significances of the CP-sensitive observables in a 
larger region of the parameter space, we now analyze the neutralino cross 
sections, branching ratios and, as an example, the asymmetry $\Aa$~\rf{eq:Obs1}  
in the $|\mu|$--$M_2$ plane. 

\begin{figure}[htbp]
  \centering
  \subfigure[]{%
    \label{fig:BRi}%
    \includegraphics[width=0.4\textwidth]{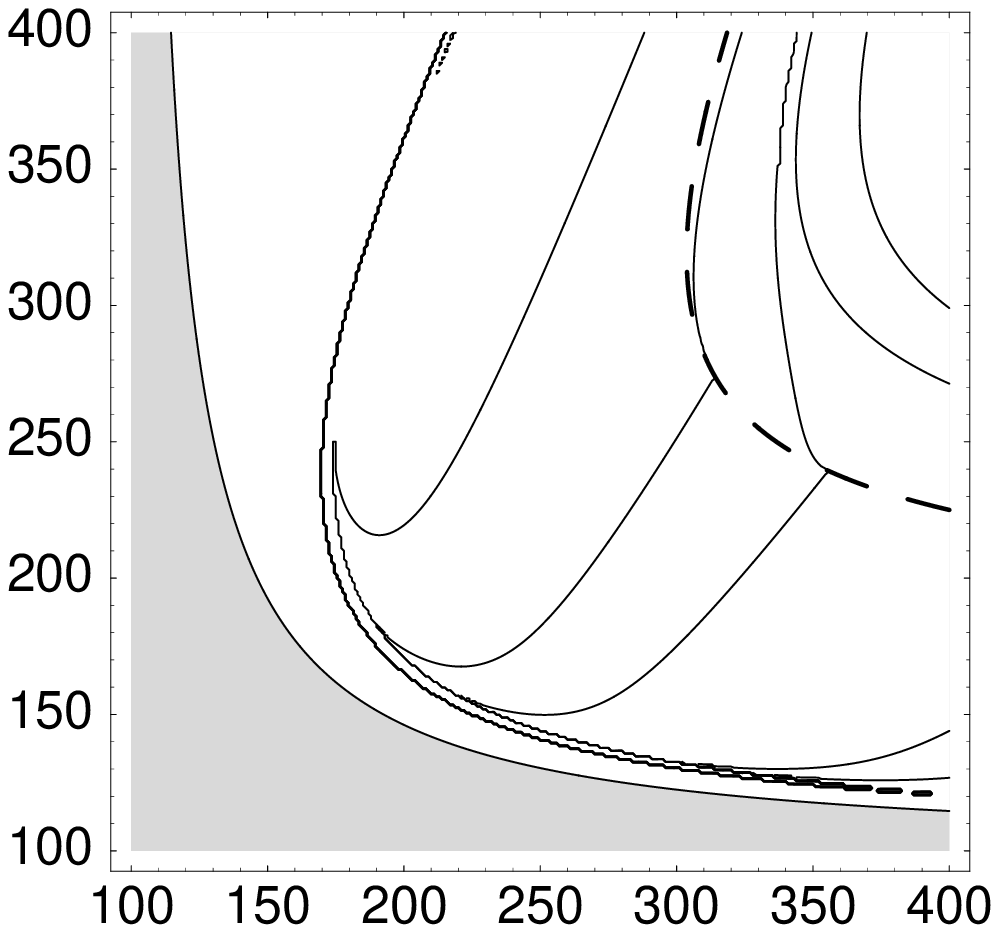}
    \put(-130,190){\fbox{${\rm BR}(\tilde\chi^0_2\to\tilde\ell_{R}\ell)$ in \%}}
    \put(-40,-15){$|\mu|$[GeV]}
    \put(-220,165){$M_2$[GeV]}
    \put(-135,150){$ A$}
    \put(-90,150){\footnotesize$66.3$}
    \put(-65,105){\footnotesize$65$}
    \put(-50,80){\footnotesize$62$}
    \put(-15,95){\footnotesize$40$}
    \put(-20,150){\footnotesize$27$}
  }
  \hspace{15mm}
  \subfigure[]{%
    \label{fig:BRj}%
    \includegraphics[width=0.4\textwidth]{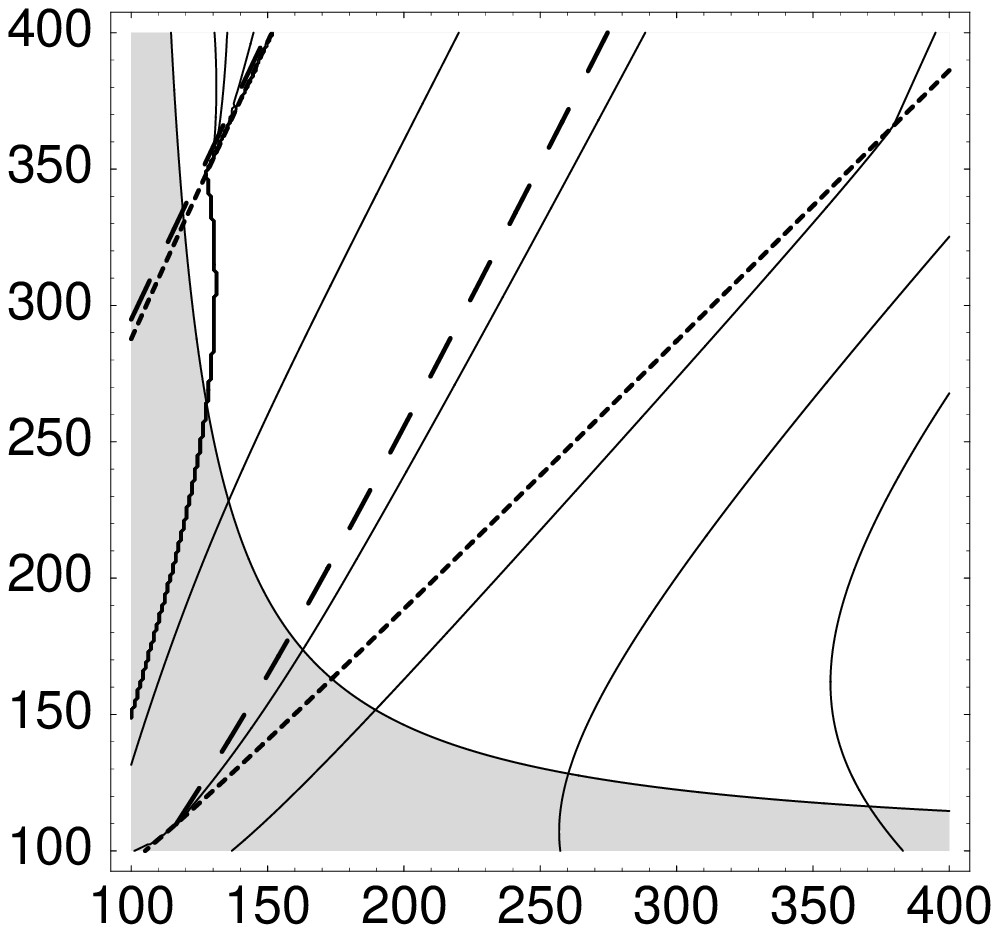}
    \put(-130,190){\fbox{${\rm BR}(\tilde\chi^0_3\to\tilde\ell_{R}\ell)$ in \%}}
    \put(-40,-15){$|\mu|$[GeV]}
    \put(-220,165){$M_2$[GeV]}
    \put(-155,150){$ A$}
    \put(-120,150){\footnotesize$66.6$}
    \put(-90,105){\footnotesize$40$}
    \put(-70,80){\footnotesize$5$}
    \put(-30,95){\footnotesize$1$}
    \put(-25,60){\footnotesize$0.5$}
  }\\ \vspace{5mm}
  \subfigure[]{%
    \label{fig:ProdXS}%
    \includegraphics[width=0.4\textwidth]{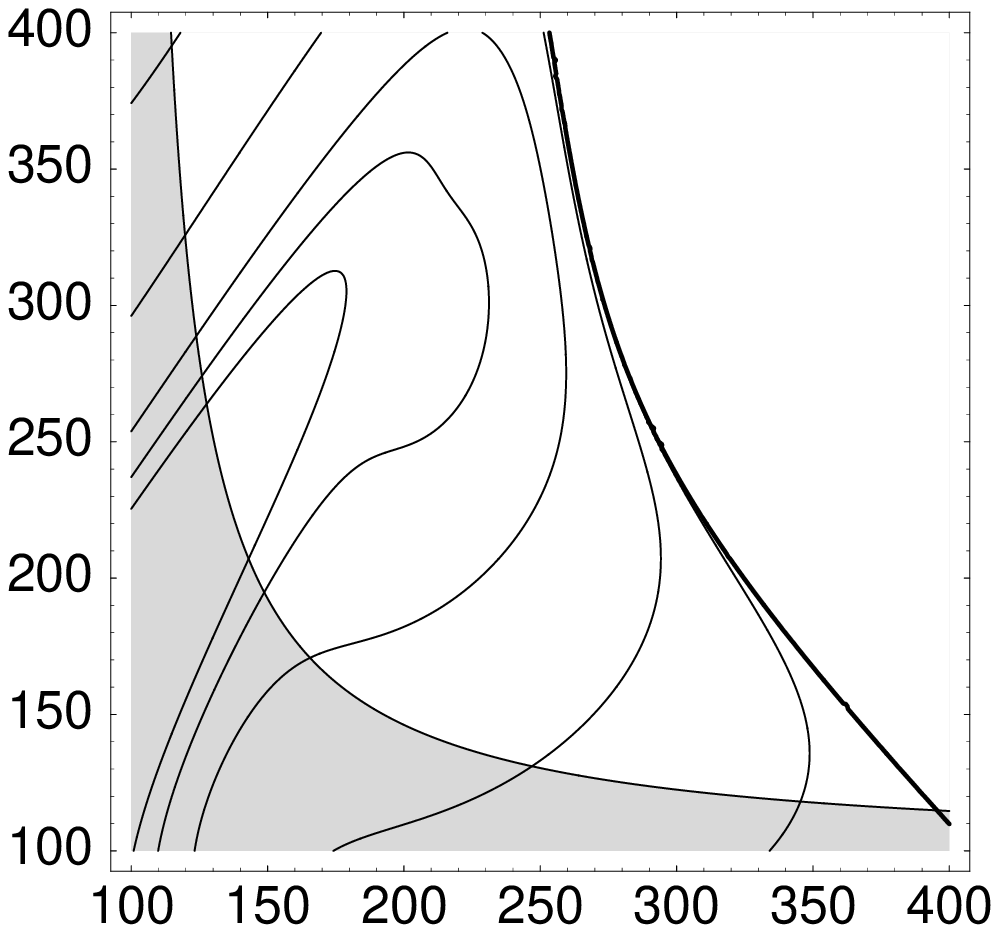}
    \put(-110,190){\fbox{$\sigma_{23}$ in fb}}
    \put(-40,-15){$|\mu|$[GeV]}
    \put(-220,165){$M_2$[GeV]}
    \put(-45,140){$ B$}
    \put(-135,100){\footnotesize$100$}
    \put(-110,95){\footnotesize$75$}
    \put(-80,80){\footnotesize$50$}
    \put(-70,35){\footnotesize$20$}
    \put(-30,30){\footnotesize$5$}
  }
  \hspace{15mm}
  \subfigure[]{%
    \label{fig:totXS}%
    \includegraphics[width=0.4\textwidth]{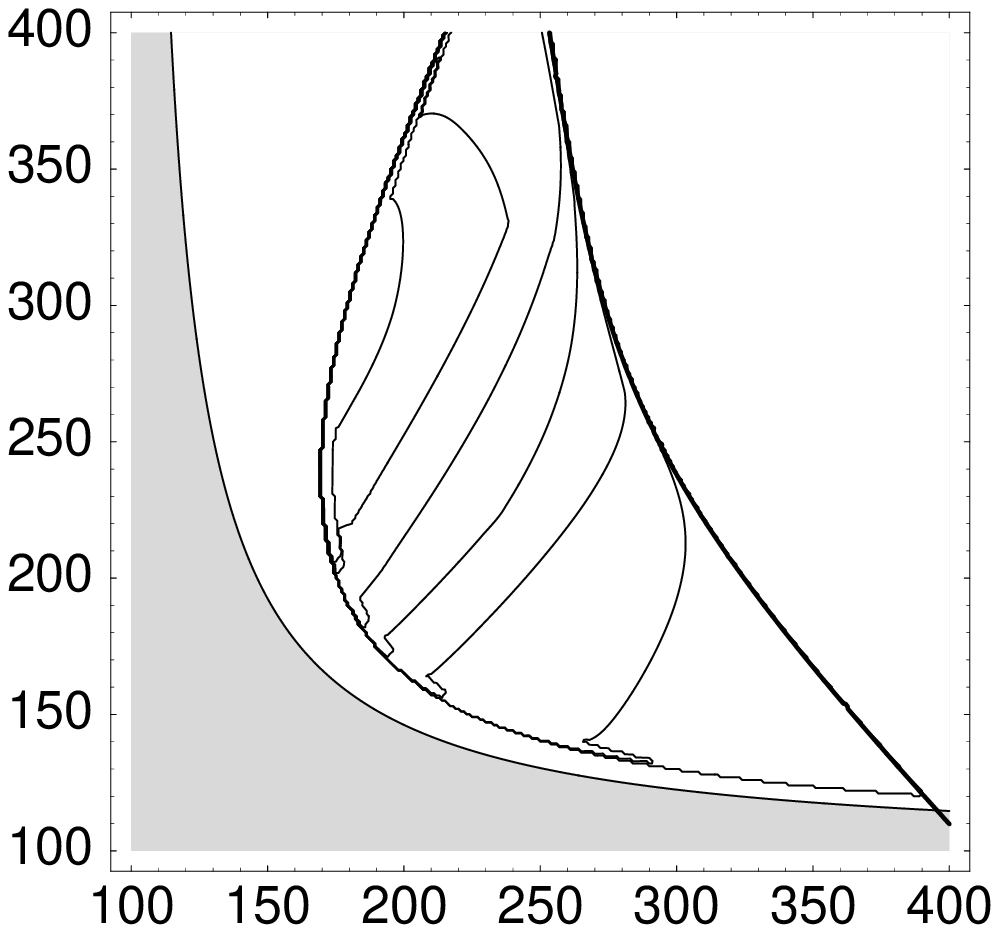}
    \put(-110,190){\fbox{$\sigma$ in fb}}
    \put(-40,-15){$|\mu|$[GeV]}
    \put(-220,165){$M_2$[GeV]}
    \put(-135,150){$ A$}
    \put(-45,140){$ B$}
    \put(-105,130){\footnotesize$37$}
    \put(-91,145){\footnotesize$30$}
    \put(-100,95){\footnotesize$10$}
    \put(-93,80){\footnotesize$4$}
    \put(-90,50){\footnotesize$1$}
    \put(-65,40){\footnotesize$0.1$}
  }
  \caption{\label{fig:BRXS}
            $|\mu|$ and $M_2$ dependence of 
           (a)~the neutralino branching ratio 
               ${\rm BR}(\tilde\chi^0_2\to\tilde\ell_{R}\ell)$,
           (b)~the branching ratio ${\rm BR}(\tilde\chi^0_3\to\tilde\ell_{R}\ell)$,
           (c)~the neutralino production cross section
               $\sigma_{23}=\sigma(e^+e^-\to\tilde\chi^0_2\tilde\chi^0_3)$, and
           (d)~the combined cross section of production and decay, 
               $\sigma=\sigma_{23}\times 
               {\rm BR}(\tilde\chi^0_2\to\tilde\ell_{R}\ell )\times
               {\rm BR}(\tilde\chi^0_3\to\tilde\ell_{R}\ell )$,  
           for the scenario as defined in Tab.~\ref{tab:scenario}.
           In region $ A$  the neutralinos are below the 
           decay threshold, $m_{\chi^0_{2,3}} < m_{\tilde\ell_R}$, and
           in region $ B$  they are above the 
           production threshold, $m_{\chi^0_2}+m_{\chi^0_3}>\sqrt{s}=500$~GeV.
           In the gray shaded areas the chargino mass is $m_{\chi^\pm_1}<100$~GeV. 
           The dashed contours in (a), (b) indicate the kinematical limit
           $m_{\chi^0_{2,3}} = m_{\chi^0_{1}} + m_Z$, respectively. 
           The dotted contour in (b) indicates the limit 
           $m_{ \chi^0_{3}} = m_W + m_{\chi^\mp_{1}}$.
          }
\end{figure}

In Figs.~\ref{fig:BRi} and~\ref{fig:BRj}, we show the neutralino branching ratios 
which are summed over both lepton flavors $\ell=e,\mu$ and charges, i.e.,
${\rm BR}(\tilde{\chi}^0_k\to\tilde\ell_{R}\ell )=  
4\times {\rm BR}(\tilde{\chi}^0_k\to\tilde e_{R}^+ e^-)$, for $k=2,3$.
In the gray shaded area, the chargino mass is $m_{\chi^\pm_1}<100$~GeV, 
and thus near or below the exclusion limit of LEP2~\cite{Amsler:2008zz}.
In region $ A$, the neutralinos are below the 
decay threshold, $m_{\chi^0_{2,3}} < m_{\tilde\ell_R}$,
and thus the corresponding two-body decays are closed.
The neutralino $\tilde\chi^0_{2}$ is always lighter than $\tilde\ell_{L}$
in the shown region of the  $|\mu|$--$M_2$ plane. We find that 
the $\tilde\chi^0_{3}$ branching ratio into left sleptons is smaller 
than ${\rm BR}(\tilde{\chi}^0_3\to\tilde\ell_{L}\ell )<1\%$.
In Fig.~\ref{fig:BRi} and \ref{fig:BRj},
the decay channels into the lightest Higgs and $Z$ bosons open
to the right of the  dashed lines, which indicate the kinematical limit 
$m_{\chi^0_{2,3}} = m_{\chi^0_{1}} + m_Z$, respectively.
However, these channels would lead to vanishing  CP observables,
due to the Majorana properties of the Higgs and $Z$ boson couplings to the 
neutralinos, as discussed in the introduction.
Along the dotted contour in Fig.~\ref{fig:BRj}, the decay channel 
$\tilde \chi^0_{3}\to W^\pm \tilde\chi^\mp_{1}$ opens, which also considerably reduces 
${\rm BR}(\tilde{\chi}^0_3\to\tilde\ell_{R}\ell )$ to the right of that contour,
for $|\mu| \gsim M_2$.  The neutralino $\tilde{\chi}^0_2$ and $\tilde{\chi}^0_3$ 
branching ratios into staus become larger than those into selectrons  for
 $|\mu| \gsim M_2$. If the tau momenta can be reconstructed,
these decay channels can also be used to measure the CP observables.
However due to stau mixing, the observables will be reduced compared
to the decays into selectrons or smuons, see the discussion 
in Ref.~\cite{Dreiner:2007ay}.

The neutralino production cross section
 $\sigma_{23}=\sigma(e^+e^-\to\tilde\chi^0_2\tilde\chi^0_3)$
is shown in Fig.~\ref{fig:ProdXS}.
It reaches values up to $130$~fb for $M_2\approx 250$~GeV and $|\mu|\approx 150$~GeV. 
In the region $ B$, the neutralinos are too heavy and above the
production threshold, $m_{\chi^0_2}+m_{\chi^0_3}>\sqrt{s}=500$~GeV.
The combined cross section of production and decay, 
$\sigma=\sigma_{23}\times 
{\rm BR}(\tilde\chi^0_2\to\tilde\ell_{R}\ell )\times
{\rm BR}(\tilde\chi^0_3\to\tilde\ell_{R}\ell )$,  
is shown in Fig.~\ref{fig:totXS}. One can see the combination of the 
kinematically excluded regions from  production and decay.
The cross section $\sigma$ reaches up to $65$~fb.

\clearpage
\begin{figure}[htbp]
  \centering
  \subfigure[]{%
    \label{fig:Asy3_M2Mu}%
    \includegraphics[width=0.4\textwidth]{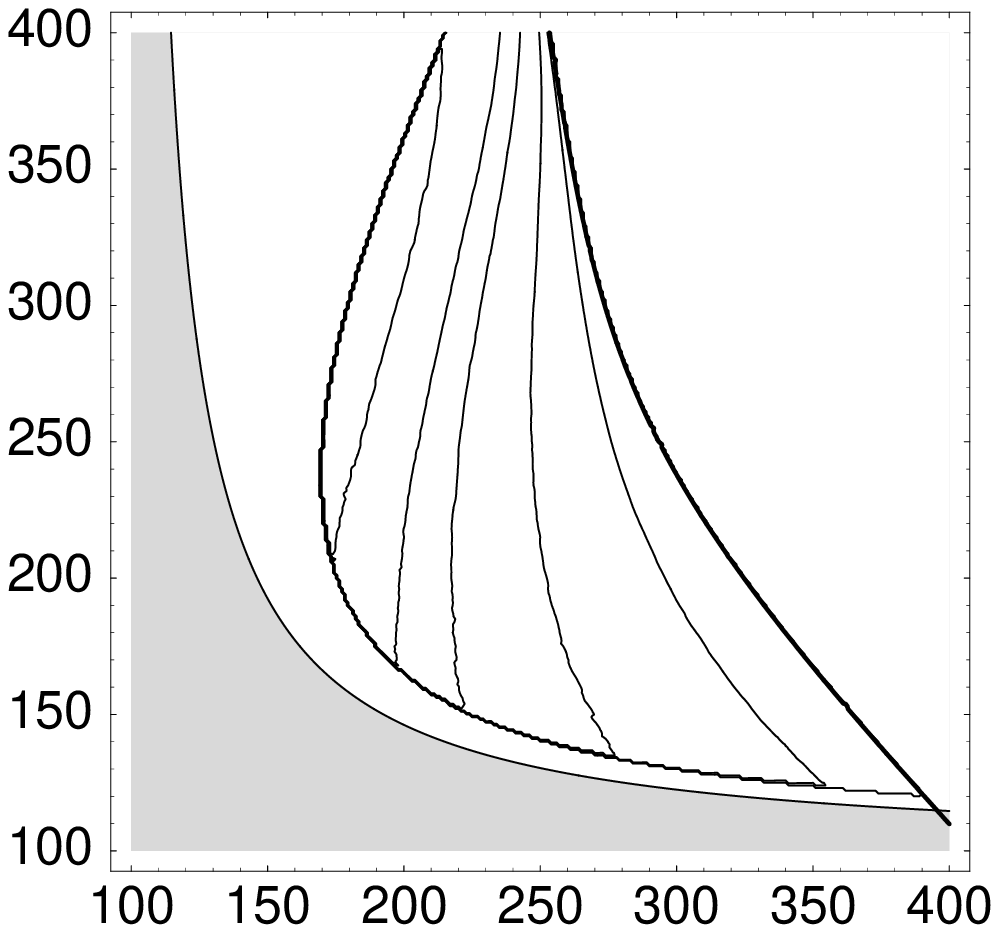}
    \put(-110,190){\fbox{$ \Aa$ in \%}}
    \put(-40,-15){$|\mu|$[GeV]}
    \put(-220,165){$M_2$[GeV]}
    \put(-135,150){$ A$}
    \put(-45,140){$B$}
    \put(-120,100){\footnotesize$-25$}
    \put(-117,65){\footnotesize$-15$}
    \put(-100,50){\footnotesize$-10$}
    \put(-85,65){\footnotesize$-5$}
    \put(-58,40){\footnotesize$-2$}
  }
  \hspace{15mm}
  \subfigure[]{%
    \label{fig:Sig3_M2Mu}%
    \includegraphics[width=0.4\textwidth]{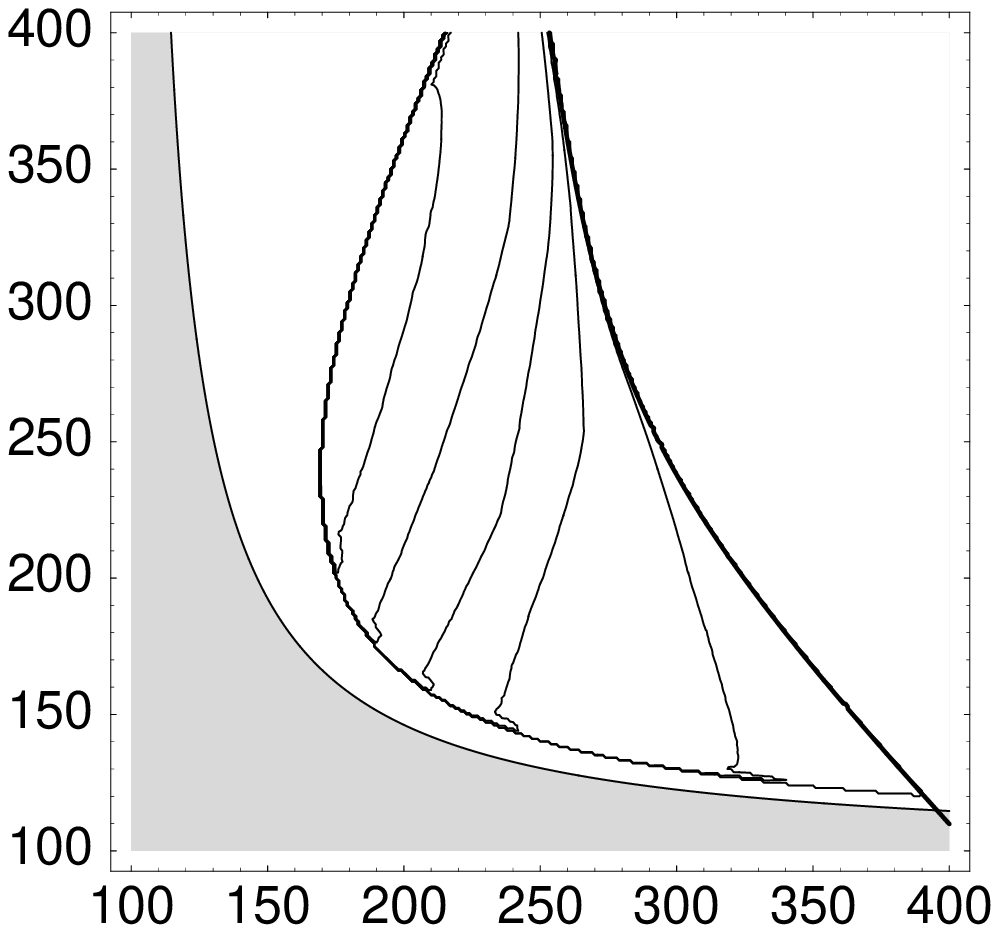}
    \put(-100,190){\fbox{$\SAa $}}
    \put(-40,-15){$|\mu|$[GeV]}
    \put(-220,165){$M_2$[GeV]}
    \put(-135,150){$ A$}
    \put(-45,140){$ B$}
    \put(-115,110){\footnotesize$30$}
    \put(-97,130){\footnotesize$10$}
    \put(-94,80){\footnotesize$3$}
    \put(-80,50){\footnotesize$1$}
    \put(-60,40){\footnotesize$0.1$}
  }
  \caption{\label{fig:AsySig3_M2Mu}
           Contour lines of 
           (a) the asymmetry $\Aa$ and 
           (b) its statistical significance $\SAa$ in the $|\mu|$--$M_2$ plane,
            for the scenario as defined in Tab.~\ref{tab:scenario}.
           In region $ A$  neutralino $\tilde\chi_2^0$  is below the 
           decay threshold, $m_{\chi^0_{2}} < m_{\tilde\ell_R}$, and
           in region $ B$  the neutralinos are above the 
           production threshold, $m_{\chi^0_2}+m_{\chi^0_3}>\sqrt{s}=500$~GeV.
           In the gray shaded areas the chargino mass is $m_{\chi^\pm_1}<100$~GeV. 
           }
\end{figure}

In Fig.~\ref{fig:AsySig3_M2Mu}, we show the asymmetry $\Aa$ and its corresponding
significance $\SAa$ in the $|\mu|$--$M_2$ plane. The asymmetry reaches values
up to $-30\%$, while the significance goes up to $50$ standard
deviations close to the kinematical limit $m_{\chi^0_2}= m_{\tilde\ell_R} $,
at $M_2\approx 300$~GeV and $|\mu|\approx 180$~GeV. At that point,
the asymmetry of the correlation  $\Tc$, that does not need the reconstruction of the
neutralino momenta, reaches $\Ac =13\%$,
which corresponds to a significance of about $\SAc =25$.


\newpage

\section{Summary and conclusions \label{conclusion}}

We have analyzed CP observables in neutralino production, which are sensitive 
to the physical phases of the gaugino parameter $M_1$, and the  higgsino parameter $\mu$. 
The observables and asymmetries rely on T-odd products in the neutralino spin-spin 
correlations, which  appear on tree-level. 
The CP-sensitive spin-spin correlations are those terms of the matrix element, 
which include the polarizations of both neutralinos, with one component normal 
to the production plane. These spin-spin correlations of the neutralinos 
can be analyzed via angular distributions of the decay leptons
$\tilde\chi_k^0 \to \tilde\ell ~ \ell$.

In order to probe the CP-sensitive spin-spin correlation terms,
we have defined different T-odd products.
One class only involves the final lepton momenta, which has the
advantage that it is not necessary to reconstruct the production plane. 
The second class of T-odd products also includes the neutralino momenta.
Based on these T-odd products, we have studied two sorts of CP-sensitive
observables. One sort are CP-sensitive observables,
which are the expectation values of the T-odd products.
The other sort are their corresponding  asymmetries, which
give the expectation value of the sign of the T-odd products.

In our numerical analysis for $\tilde\chi_2^0\tilde\chi_3^0$ production, 
we have found that the observables are
largest in mixed scenarios with small $\tan\beta$. We have defined theoretical 
significances to decide, which CP observable is most sensitive to the
CP phases.
For a linear collider with $\sqrt s=500$~GeV and 
longitudinally polarized beams,
$({\mathcal P}_-,{\mathcal P}_+)=(0.9,-0.6)$,
with an integrated luminosity of ${\mathcal L}=500~{\rm fb}^{-1}$,
the CP-sensitive observables that only include the momenta of the decay 
leptons yield theoretical significances of $S\lsim 25$ for $\phi_1=0.5\pi$.
We find larger theoretical significances up to $S\lsim 50$ for the CP-sensitive observables 
that need a reconstruction of the neutralino momenta.
However, only a detailed experimental study with
background and detector simulations can show whether the CP-sensitive 
observables are accessible.
We hope that our results motivate such a study.

\section*{Acknowledgments}

M.~T. thanks R.~K\"ogerler for very useful discussions and encouragement.
This work is supported by the ``Fonds zur F\"orderung der
wissenschaftlichen Forschung'' (FWF) of Austria, project No. P18959-N16, and
by MICINN project FPA.2006-05294.
The authors acknowledge support from EU under the MRTN-CT-2006-035505
and MTRN-CT-2006-503369 network proprammes. 
T.~K. is supported by the Portuguese FCT through the projects
POCI/FP/81919/2007 and CFTP-FCT UNIT 777, which are partially
funded through POCTI (FEDER).
%
%
\begin{appendix}
\setcounter{equation}{0}
\renewcommand{\thesubsection}{\Alph{section}.\arabic{subsection}}
\renewcommand{\theequation}{\Alph{section}.\arabic{equation}}
\section*{Appendix}
%
\section{Decay terms $D$ and $\Sigma^{c}_{D}$ \label{Decaydensity}}

The coefficients in Eq.~\rf{eq:amplitudesquared} of the neutralino 
decay matrices for  the decay into right sleptons
$\tilde\chi_k^0 \to \tilde\ell_{R}^-~\ell^+$, 
with  $\ell = e,\mu$, are~\cite{Kittel:2004rp}
\begin{eqnarray}
D_k &=& \phantom{+}
\frac{g^2}{2} |f^{R}_{\ell k}|^2 (m_{\chi_k^0}^2 -m_{\tilde\ell_R}^2 )~,\\[2mm]
\Sigma^c_{D_k} &=&  \,^{\;\,+}_{(-)} g^2 |f^{R}_{\ell k}|^2 
m_{\chi_k^0} (s^c_{\chi_k} \cdot p_{\ell})~,
\label{neutdecay1}
\end{eqnarray}
where the sign in parenthesis 
holds for the charge conjugated process 
$\tilde\chi_k^0 \to \tilde\ell_{R}^+~\ell^-$.
\\
For the decay into the left sleptons
$\tilde\chi^0_k \to \tilde\ell_L^-~\ell^+$, $\ell=e,\mu$,
the coefficients are
\begin{eqnarray}
D_k &=& \phantom{-}
       \frac{g^2}{2} |f^{L}_{\ell k}|^2 
                (m_{\chi_k^0}^2 - m_{\tilde\ell_L}^2 )~,\\[2mm]
\Sigma^c_{D_k} &=&  \,^{\;\,-}_{(+)} g^2 |f^{L}_{\ell k}|^2 m_{\chi_k^0} 
                (s^c_{\chi_k} \cdot p_{\ell})~,
\label{neutdecay2}
\end{eqnarray}
where the sign in parenthesis 
holds for the charge conjugated process 
$\tilde\chi^0_k \to \tilde\ell_L^+~\ell^-$.
%

In order to reduce the free MSSM parameters, we parametrize the slepton masses with
an approximate solution to the renormalization group equations (RGE)~\cite{RGE}
\begin{eqnarray}
        m_{\tilde\ell_R}^2 &=& m_0^2 +m_\ell^2+0.23 M_2^2
        -m_Z^2\cos 2 \beta \sin^2 \theta_W~,\label{mslr}\\ 
        m_{\tilde\ell_L  }^2 &=& m_0^2 +m_\ell^2+0.79 M_2^2
        +m_Z^2\cos 2 \beta(-\frac{1}{2}+ \sin^2 \theta_W)~,\label{msll} \\
       m_{\tilde\nu_{\ell}  }^2 &=& m_0^2 +m_\ell^2+0.79 M_2^2 +
       \frac{1}{2}m_Z^2\cos 2 \beta~,
\end{eqnarray}
with $m_0$ the common scalar mass parameter at the GUT scale.

\section{Momentum and polarization vectors\label{Vectors}}

We choose a coordinate system with the $z$-axis
along the $\vec{p}_{e^-}$ direction in the center-of-mass system.
The four-momenta of the neutralinos $\ti\chi^0_i$ and $\ti\chi^0_j$ are 
\baq{eq:momentumchar}
 p_{\chi_{i,j}}&=&q~
(\frac{E_{\chi_{i,j}}}{q},\mp\sin\theta,0,\mp\cos\theta)~,
\eaq
with their energies and common momentum
\be{eq:energy}
E_{\chi_{i,j}}=\frac{s+m^2_{\chi_{i,j}}-m^2_{\chi_{j,i}}}{2 \sqrt{s}}~,\qquad
q=\frac{\lambda^{\frac{1}{2}}(s,m^2_{\chi_i},m^2_{\chi_j})}{2 \sqrt{s}}~,
\ee
respectively, and the kinematic function 
$\lambda(a,b,c)=a^2+b^2+c^2-2(a b + a c + b c)$.
The scattering angle is $\theta \varangle (\vec{p}_{e^-},\vec{p}_{\chi_j})$, 
whereas the azimuthal angle can be set to zero, due to rotational 
invariance around the beam axis~\cite{Byckling}.

The three spin basis vectors of $\ti\chi^0_i$ and $\ti\chi^0_j$
are chosen to be 
\baq{eq:polvec}
s^1_{\chi_{i,j}}&=&\left(0,\frac{\vec{s}_{\chi_{i,j}}^{\,2}\times\vec{s}_{\chi_{i,j}}^{\,3}}
{|\vec{s}_{\chi_{i,j}}^{\,2}\times\vec{s}_{\chi_{i,j}}^{\,3}|}\right)=
\pm(0, \cos\theta,0,-\sin\theta)~,
\nonumber \\[3mm]
s^2_{\chi_{i,j}}&=&\left(0,
\frac{\vec{p}_{e^-}\times \vec{p}_{\chi_{i,j}}}
{|\vec{p}_{e^-}\times\vec{p}_{\chi_{i,j}}|}\right)=
(0,0,1,0)~,
\nonumber \\[3mm]
s^3_{\chi_{i,j}}&=&\frac{1}{m_{\chi_{i,j}}}
\left(q, 
\frac{E_{\chi_{i,j}}}{q}~\vec{p}_{\chi_{i,j}} \right)=
\frac{E_{\chi_{i,j}}}{m_{\chi_{i,j}}}
\left(\frac{ q}{E_{\chi_{i,j}}  },\mp \sin\theta,0,\mp\cos\theta\right)~.
\eaq
They fulfill the orthonormality relations
$s_{\chi_k}^{c}\cdot s_{\chi_k}^{d}=-\delta^{cd}$,
$s_{\chi_k}^{c}\cdot p_{\chi_k}=0$,
and the completeness relation~\cite{Moortgat-Pick:1999di,Haber:1994pe}  
\baq{eq:chicompleteness}
\sum_c s_{\chi_k}^{c,\,\mu} \cdot s_{\chi_k}^{c,\,\nu}=
 -g^{\mu\nu}+\frac{p_{\chi_k}^\mu p_{\chi_k}^\nu}{ m_{\chi_k}^2}~. 
\eaq
The four-momenta of the leptons in the decays 
$\tilde\chi_i^0 \to \tilde\ell~\ell$, and
$\tilde\chi_j^0 \to \tilde\ell'~\ell'$, are
\be{eq:fourlp}
p_{\ell}=
|\vec{p}_{\ell}| (1,\cos\phi_{\ell}
\sin\theta_{\ell},
\sin\phi_{\ell} \sin\theta_{\ell},
\cos\theta_{\ell})~,
\ee
\be{eq:fourlm}
p_{\ell'}=
|\vec{p}_{\ell'}| (1,\cos\phi_{\ell'}
\sin\theta_{\ell'},
\sin\phi_{\ell'} \sin\theta_{\ell'},
\cos\theta_{\ell'})~,
\ee
respectively, with
\be{eq:lepmom}
|\vec{p}_{\ell}|=\frac{m^2_{\chi_i}-
m^2_{\tilde\ell}}{2(E_{\chi_i}+q\cos\vartheta_{\ell})}~,
\quad
|\vec{p}_{\ell'}|=\frac{m^2_{\chi_j}-
m^2_{\tilde\ell'}}{2(E_{\chi_j}-q\cos\vartheta_{\ell'})}~,
\ee
and the decay angles
\baq{eq:angle}
\cos\vartheta_{\ell}&=&\sin\theta \sin\theta_{\ell} \cos\phi_{\ell}+
\cos\theta \cos\theta_{\ell}~,
\nonumber\\[2mm]
\cos\vartheta_{\ell'}&=&\sin\theta \sin\theta_{\ell'} \cos\phi_{\ell'}+
\cos\theta \cos\theta_{\ell'}~.
\eaq
With these definitions, the T-odd products $f^{ab}$~\rf{eq:f8} 
of the spin-spin correlation terms  in the laboratory system are
\begin{eqnarray}
f^{12} &=& -\frac{1}{2}E_{\chi_i} s q \sin^2\theta~,
\qquad
f^{21} = \frac{1}{2}E_{\chi_j} s q \sin^2\theta~,
\label{eq:f21}\\[2mm]
f^{23} &=& \frac{1}{4} m_{\chi_j} s q \sin(2\theta)~,
\qquad
f^{32} = -\frac{1}{4} m_{\chi_i} s q \sin(2\theta)~.
\label{eq:f32}
\end{eqnarray}
%

\section{Phase space \label{Phase space}}

The Lorentz invariant phase space element in Eq.~\rf{eq:crossection}
is given by~\cite{Kittel:2004rp,Byckling}
\baq{eq:phasespace}
{\rm d Lips} =\frac{1}{(2 \pi)^2} {\rm d Lips}(s,p_{\chi_i},p_{\chi_j})
{\rm d}s_{\chi_i}
{\rm d Lips}(s_{\chi_i},p_{\ti\ell},p_{\ell})
{\rm d}s_{\chi_j}
{\rm d Lips}(s_{\chi_j},p_{\ti\ell'},p_{\ell'}),
\nonumber\\
\eaq
with $s_{\chi_{i,j}}=p^2_{\chi_{i,j}}$.
The different factors of the phase space element are
\be{eq:prodphs}
{\rm d Lips}(s,p_{\chi_i},p_{\chi_j})=\frac{1}{8 \pi} \frac{q}{\sqrt{s}}
\sin\theta~ {\rm d}\theta~,
\ee
\be{eq:lepphsp}
{\rm d Lips}(s_{\chi_i},p_{\ti\ell},p_{\ell})=
\frac{1}{2(2 \pi)^2}\frac{|\vec{p}_{\ell}|^2}{m^2_{\chi_i}-m^2_{\ti\ell}}
\sin\theta_{\ell}~ {\rm d}\theta_{\ell}~ {\rm d}\phi_{\ell}~,
\ee
%
%
\be{eq:lepphsm}
{\rm d Lips}(s_{\chi_j},p_{\ti\ell'},p_{\ell'})=
\frac{1}{2(2 \pi)^2}\frac{|\vec{p}_{\ell'}|^2}{m^2_{\chi_j}-m^2_{\ti\ell'}}
\sin\theta_{\ell'}~ {\rm d}\theta_{\ell'}~ {\rm d}\phi_{\ell'}~.
\ee
We use the narrow width approximation for the propagators
in Eq.~\rf{eq:amplitudesquared},
\be{eq:narrowwidth}
\int|\Delta(\CH_{i,j}^0)|^2  {\rm d} s_{\chi_{i,j}} = 
\frac{\pi}{m_{\chi_{i,j}}\Gamma_{\chi_{i,j}}}~,
\ee
which is justified for $\Gamma/m\ll1$, which holds in our case with
$\Gamma\lsim {\mathcal O}(1 {\rm GeV}) $. Note, however, that the naive
${\mathcal O}(\Gamma/m)$-expectation of the error can easily receive
large off-shell corrections of an order of magnitude and more,
in particular at threshold, or due to interferences
with other resonant or non-resonant processes.
For recent discussions of these issues, see Ref.~\cite{narrowwidth}.

\end{appendix}




\begin{thebibliography}{99}


\bibitem{Gavela:1993ts}
  M.~B.~Gavela, P.~Hernandez, J.~Orloff and O.~Pene,
  Mod.\ Phys.\ Lett.\  A {\bf 9}, 795 (1994)
  [arXiv:hep-ph/9312215];\\
%
  M.~B.~Gavela, P.~Hernandez, J.~Orloff, O.~Pene and C.~Quimbay,
  Nucl.\ Phys.\  B {\bf 430}, 382 (1994)
  [arXiv:hep-ph/9406289];\\
%
  F.~Csikor, Z.~Fodor and J.~Heitger,
  Phys.\ Rev.\ Lett.\  {\bf 82} (1999) 21
  [arXiv:hep-ph/9809291].

\bibitem{Riotto:1998bt}
  A.~Riotto,
  arXiv:hep-ph/9807454;
%
  W.~Bernreuther,
  Lect.\ Notes Phys.\  {\bf 591} (2002) 237
  [arXiv:hep-ph/0205279].

\bibitem{Ellis:2008zy}
  J.~R.~Ellis, J.~S.~Lee and A.~Pilaftsis,
  JHEP {\bf 0810} (2008) 049
  [arXiv:0808.1819 [hep-ph]].

\bibitem{EDM}
  F.~del Aguila, M.~B.~Gavela, J.~A.~Grifols and A.~Mendez,
  Phys.\ Lett.\  B {\bf 126}, 71 (1983)
  [Erratum-ibid.\  B {\bf 129}, 473 (1983)];\\
%
%
  Y.~Kizukuri and N.~Oshimo,
  Phys.\ Rev.\  D {\bf 46}, 3025 (1992);\\
%
  T.~Ibrahim and P.~Nath,
  Phys.\ Rev.\ D {\bf 57} (1998) 478
  [Erratum-ibid.\ D {\bf 58} (1998) 019901, D {\bf 60} (1999) 079903,
  D {\bf 60} (1999) 119901] 
  [arXiv:hep-ph/9708456];\\
%
  M.~Brhlik, G.~J.~Good and G.~L.~Kane,
  Phys.\ Rev.\ D {\bf 59}, 115004 (1999)
  [arXiv:hep-ph/9810457];\\
%
  A.~Bartl, T.~Gajdosik, W.~Porod, P.~Stockinger and H.~Stremnitzer,
  Phys.\ Rev.\  D {\bf 60} (1999) 073003
  [arXiv:hep-ph/9903402];\\
%
  D.~Chang, W.~Y.~Keung and A.~Pilaftsis,
  Phys.\ Rev.\ Lett.\  {\bf 82}, 900 (1999)
  [Erratum-ibid.\  {\bf 83}, 3972 (1999)]
  [arXiv:hep-ph/9811202];\\
%
%
  V.~D.~Barger, T.~Falk, T.~Han, J.~Jiang, T.~Li and T.~Plehn,
  Phys.\ Rev.\ D {\bf 64}, 056007 (2001)
  [arXiv:hep-ph/0101106];\\
%
  S.~Abel, S.~Khalil and O.~Lebedev,
  Nucl.\ Phys.\  B {\bf 606}, 151 (2001)
  [arXiv:hep-ph/0103320];\\
%
  S.~Yaser Ayazi and Y.~Farzan,
  Phys.\ Rev.\ D {\bf 74} (2006) 055008
  [arXiv:hep-ph/0605272].


\bibitem{ILC}
  J.~Brau {\it et al.}  [ILC Collaboration],
  arXiv:0712.1950 [physics.acc-ph];\\
%
  J.~A.~Aguilar-Saavedra {\it et al.}  [ECFA/DESY LC Physics Working Group],
  arXiv:hep-ph/0106315;\\
%
  T.~Abe {\it et al.}  [American Linear Collider Working Group],
  arXiv:hep-ex/0106055;\\
%
  K.~Abe {\it et al.}  [ACFA Linear Collider Working Group],
  arXiv:hep-ph/0109166;\\
  J.~A.~Aguilar-Saavedra {\it et al.},
  Eur.\ Phys.\ J.\ C {\bf 46} (2006) 43
  [arXiv:hep-ph/0511344].
%



\bibitem{Moortgat-Pick:1999di}
  G.~A.~Moortgat-Pick, H.~Fraas, A.~Bartl and W.~Majerotto,
  Eur.\ Phys.\ J.\  C {\bf 9} (1999) 521
  [Erratum-ibid.\  C {\bf 9} (1999) 549]
  [arXiv:hep-ph/9903220];\\
%
  G.~A.~Moortgat-Pick, Doctoral thesis
``Spin effects in chargino/neutralino production and decay'' (in German),
Universit\"at W\"urzburg (1999). 


\bibitem{Kneur:1999nx}
J.~L.~Kneur and G.~Moultaka,
Phys.\ Rev.\ D {\bf 61} (2000) 095003
[arXiv:hep-ph/9907360];\\
%
V.~D.~Barger, T.~Han, T.~J.~Li and T.~Plehn,
Phys.\ Lett.\ B {\bf 475} (2000) 342
[arXiv:hep-ph/9907425];\\
%
  S.~Y.~Choi, H.~S.~Song and W.~Y.~Song,
  Phys.\ Rev.\  D {\bf 61} (2000) 075004
  [arXiv:hep-ph/9907474];\\
%
  S.~Y.~Choi, J.~Kalinowski, G.~A.~Moortgat-Pick and P.~M.~Zerwas,
  Eur.\ Phys.\ J.\  C {\bf 22} (2001) 563
  [Addendum-ibid.\  C {\bf 23} (2002) 769]
  [arXiv:hep-ph/0108117];\\
%
G.~J.~Gounaris and C.~Le Mouel,
Phys.\ Rev.\ D {\bf 66} (2002) 055007
[arXiv:hep-ph/0204152];\\
%
S.~Y.~Choi,
Phys.\ Rev.\ D {\bf 69} (2004) 096003
[arXiv:hep-ph/0308060].


\bibitem{Valencia:1994zi}
G.~Valencia,
arXiv:hep-ph/9411441, and references therein;\\
G.C.~Branco, L.~Lavoura, and J.P.~Silva, {\em CP violation}, 
Oxford University Press, New York, 1999.

\bibitem{Dreiner:2007ay}
  H.~K.~Dreiner, O.~Kittel and F.~von der Pahlen,
  JHEP {\bf 0801}, 017 (2008)
  [arXiv:0711.2253 [hep-ph]].


\bibitem{Bartl:2008fu}
  A.~Bartl, K.~Hohenwarter-Sodek, T.~Kernreiter, O.~Kittel and M.~Terwort,
  Nucl.\ Phys.\  B {\bf 802} (2008) 77
  [arXiv:0802.3592 [hep-ph]].


\bibitem{neut}
Y.~Kizukuri and N.~Oshimo,
Phys.\ Lett.\ B {\bf 249} (1990) 449;\\
%
  A.~Bartl, H.~Fraas, O.~Kittel and W.~Majerotto,
  Phys.\ Rev.\  D {\bf 69} (2004) 035007
  [arXiv:hep-ph/0308141];  
  Eur.\ Phys.\ J.\  C {\bf 36} (2004) 233
  [arXiv:hep-ph/0402016];\\
%
  A.~Bartl, T.~Kernreiter and O.~Kittel,
  Phys.\ Lett.\  B {\bf 578}, 341 (2004)
  [arXiv:hep-ph/0309340]; \\
%
  S.~Y.~Choi, M.~Drees, B.~Gaissmaier and J.~Song,
  Phys.\ Rev.\  D {\bf 69}, 035008 (2004)
  [arXiv:hep-ph/0310284];\\
%
  S.~Y.~Choi and Y.~G.~Kim,
  Phys.\ Rev.\  D {\bf 69}, 015011 (2004)
  [arXiv:hep-ph/0311037];\\
%
  J.~A.~Aguilar-Saavedra,
  Nucl.\ Phys.\  B {\bf 697} (2004) 207
  [arXiv:hep-ph/0404104];\\
%
  S.~Y.~Choi, M.~Drees and B.~Gaissmaier,
  Phys.\ Rev.\  D {\bf 70} (2004) 014010
  [arXiv:hep-ph/0403054];\\
%
  A.~Bartl, H.~Fraas, S.~Hesselbach, K.~Hohenwarter-Sodek and G.~A.~Moortgat-Pick,
  JHEP {\bf 0408} (2004) 038
  [arXiv:hep-ph/0406190];\\
%
  S.~Y.~Choi, B.~C.~Chung, J.~Kalinowski, Y.~G.~Kim and K.~Rolbiecki,
  Eur.\ Phys.\ J.\  C {\bf 46}, 511 (2006)
  [arXiv:hep-ph/0504122].



\bibitem{Kittel:2004rp}
  O.~Kittel,
  arXiv:hep-ph/0504183.



\bibitem{Bartl:2005uh}
  A.~Bartl, H.~Fraas, S.~Hesselbach, K.~Hohenwarter-Sodek, 
  T.~Kernreiter and G.~A.~Moortgat-Pick,
  JHEP {\bf 0601}, 170 (2006)
  [arXiv:hep-ph/0510029].


\bibitem{Trans}
  G.~A.~Moortgat-Pick {\it et al.},
  Phys.\ Rept.\  {\bf 460}, 131 (2008)
  [arXiv:hep-ph/0507011];\\
%
  S.~Y.~Choi, M.~Drees and J.~Song,
  JHEP {\bf 0609}, 064 (2006)
  [arXiv:hep-ph/0602131];\\
%
  A.~Bartl, K.~Hohenwarter-Sodek, T.~Kernreiter and O.~Kittel,
  JHEP {\bf 0709}, 079 (2007)
  [arXiv:0706.3822 [hep-ph]].





\bibitem{Bartl:2003ck}
  A.~Bartl, H.~Fraas, T.~Kernreiter and O.~Kittel,
  Eur.\ Phys.\ J.\  C {\bf 33}, 433 (2004)
  [arXiv:hep-ph/0306304]; \\
  J.~A.~Aguilar-Saavedra,
  Phys.\ Lett.\  B {\bf 596}, 247 (2004)
  [arXiv:hep-ph/0403243];\\
%
  P.~Langacker, G.~Paz, L.~T.~Wang and I.~Yavin,
  JHEP {\bf 0707}, 055 (2007)
  [arXiv:hep-ph/0702068];\\
  J.~Ellis, F.~Moortgat, G.~Moortgat-Pick, J.~M.~Smillie and J.~Tattersall,
  arXiv:0809.1607 [hep-ph].


\bibitem{Bartl:1986hp}
  A.~Bartl, H.~Fraas and W.~Majerotto,
  Nucl.\ Phys.\  B {\bf 278} (1986) 1.

\bibitem{haberkane}
  H.~E.~Haber and G.~L.~Kane,
  Phys.\ Rept.\  {\bf 117} (1985) 75.

\bibitem{Haber:1994pe}
  H.~E.~Haber,
  \textit{Proceedings of the 21st
          SLAC Summer Institute on Particle Physics},
          eds. L. DeProcel, Ch. Dunwoodie, Stanford 1993, 231,
  arXiv:hep-ph/9405376;\\
  H.~K.~Dreiner, H.~E.~Haber and S.~P.~Martin,
  arXiv:0812.1594 [hep-ph].





\bibitem{optimal} 
  D.~Atwood and A.~Soni,
  Phys.\ Rev.\  D {\bf 45}, 2405 (1992);\\
  M.~Diehl and O.~Nachtmann,
  Z.\ Phys.\  C {\bf 62}, 397 (1994);\\
  B.~Grzadkowski and J.~F.~Gunion,
  Phys.\ Lett.\  B {\bf 350}, 218 (1995)
  [arXiv:hep-ph/9501339].

\bibitem{Bartl:2006bn}
  A.~Bartl, H.~Fraas, K.~Hohenwarter-Sodek, T.~Kernreiter, G.~Moortgat-Pick and A.~Wagner,
  Phys.\ Lett.\  B {\bf 644}, 165 (2007)
  [arXiv:hep-ph/0610431].


\bibitem{CHAR}
  Y.~Kizukuri and N.~Oshimo,
  arXiv:hep-ph/9310224;\\
%
  S.~Y.~Choi, A.~Djouadi, M.~Guchait, J.~Kalinowski, H.~S.~Song and P.~M.~Zerwas,
  Eur.\ Phys.\ J.\ C {\bf 14}, 535 (2000)
  [arXiv:hep-ph/0002033];\\
%
  A.~Bartl, H.~Fraas, O.~Kittel and W.~Majerotto,
  Phys.\ Lett.\  B {\bf 598}, 76 (2004)
  [arXiv:hep-ph/0406309];\\
%
  O.~Kittel, A.~Bartl, H.~Fraas and W.~Majerotto,
  Phys.\ Rev.\  D {\bf 70}, 115005 (2004)
  [arXiv:hep-ph/0410054];\\
%
  J.~A.~Aguilar-Saavedra,
  Nucl.\ Phys.\  B {\bf 717}, 119 (2005)
  [arXiv:hep-ph/0410068];\\
%
%
  A.~Bartl, H.~Fraas, S.~Hesselbach, K.~Hohenwarter-Sodek, T.~Kernreiter and G.~Moortgat-Pick,
  Eur.\ Phys.\ J.\  C {\bf 51}, 149 (2007)
  [arXiv:hep-ph/0608065];\\
%
  A.~Bartl, K.~Hohenwarter-Sodek, T.~Kernreiter and H.~Rud,
  Eur.\ Phys.\ J.\ C {\bf 36} (2004) 515
  [arXiv:hep-ph/0403265].


\bibitem{Oller:2005xg}
  W.~Oller, H.~Eberl and W.~Majerotto,
  Phys.\ Rev.\  D {\bf 71} (2005) 115002
  [arXiv:hep-ph/0504109];
  Phys.\ Lett.\  B {\bf 590} (2004) 273
  [arXiv:hep-ph/0402134];\\
%
  T.~Fritzsche and W.~Hollik,
  Nucl.\ Phys.\ Proc.\ Suppl.\  {\bf 135}, 102 (2004)
  [arXiv:hep-ph/0407095].


\bibitem{Drees:2006um}
  M.~Drees, W.~Hollik and Q.~Xu,
  JHEP {\bf 0702} (2007) 032
  [arXiv:hep-ph/0610267].

\bibitem{Osland:2007xw}
  P.~Osland and A.~Vereshagin,
  Phys.\ Rev.\  D {\bf 76}, 036001 (2007)
  [arXiv:0704.2165 [hep-ph]];\\
%
  K.~Rolbiecki and J.~Kalinowski,
  Phys.\ Rev.\  D {\bf 76}, 115006 (2007)
  [arXiv:0709.2994 [hep-ph]].


\bibitem{Amsler:2008zz}
  C.~Amsler et al.\  [Particle Data Group],
  {\em Phys. Lett.}  {\bf B 667} (2008) 1.

\bibitem{RGE}
  L.~E.~Ibanez and C.~Lopez,
  Nucl.\ Phys.\ B {\bf 233} (1984) 511;\\
%
  L.~E.~Ibanez, C.~Lopez and C.~Munoz,
  Nucl.\ Phys.\ B {\bf 256} (1985) 218;\\
%
L.~J.~Hall and J.~Polchinski,
Phys.\ Lett.\ B {\bf 152} (1985) 335.


\bibitem{Byckling}  E. Byckling, K. Kajantie, {\it Particle Kinematics},
                   John Wiley \& Sons, 1973.


\bibitem{narrowwidth}
  K.~Hagiwara {\it et al.},
  Phys.\ Rev.\  D {\bf 73} (2006) 055005
  [arXiv:hep-ph/0512260];\\
%
  D.~Berdine, N.~Kauer and D.~Rainwater,
  Phys.\ Rev.\ Lett.\  {\bf 99}, 111601 (2007)
  [arXiv:hep-ph/0703058]; \\
  N.~Kauer,
  Phys.\ Lett.\  B {\bf 649}, 413 (2007)
  [arXiv:hep-ph/0703077];
  JHEP {\bf 0804}, 055 (2008)
  [arXiv:0708.1161 [hep-ph]];\\
  C.~F.~Uhlemann and N.~Kauer,
  Nucl.\ Phys.\  B {\bf 814}, 195 (2009)
  [arXiv:0807.4112 [hep-ph]];\\
  M.~A.~Gigg and P.~Richardson,
  arXiv:0805.3037 [hep-ph].




\end{thebibliography}
\end{document}